\documentclass[journal]{IEEEtran}
\usepackage{algorithmic}
\usepackage{algorithm}
\usepackage{array}
\usepackage{textcomp}
\usepackage{stfloats}
\usepackage{url}
\usepackage{verbatim}
\usepackage{graphicx}
\usepackage{cite}
\usepackage{comment, float, balance, tabularx, pifont}
\usepackage{subfig}
\usepackage{tikz}
\usepackage[colorinlistoftodos,textsize=small]{todonotes}
\usepackage{acronym}

\usepackage{scalerel}
\usepackage{amsmath}
\usepackage{amsfonts}
\usepackage{siunitx}  
\usepackage{tikz}
\usepackage{booktabs}
\usepackage{multirow}
\usepackage{xcolor}
\usetikzlibrary{svg.path}

\definecolor{orcidlogocol}{HTML}{A6CE39}
\tikzset{
  orcidlogo/.pic={
    \fill[orcidlogocol] svg{M256,128c0,70.7-57.3,128-128,128C57.3,256,0,198.7,0,128C0,57.3,57.3,0,128,0C198.7,0,256,57.3,256,128z};
    \fill[white] svg{M86.3,186.2H70.9V79.1h15.4v48.4V186.2z}
                 svg{M108.9,79.1h41.6c39.6,0,57,28.3,57,53.6c0,27.5-21.5,53.6-56.8,53.6h-41.8V79.1z M124.3,172.4h24.5c34.9,0,42.9-26.5,42.9-39.7c0-21.5-13.7-39.7-43.7-39.7h-23.7V172.4z}
                 svg{M88.7,56.8c0,5.5-4.5,10.1-10.1,10.1c-5.6,0-10.1-4.6-10.1-10.1c0-5.6,4.5-10.1,10.1-10.1C84.2,46.7,88.7,51.3,88.7,56.8z};
  }
}

\newcommand\orcidicon[1]{\href{https://orcid.org/#1}{\mbox{\scalerel*{
\begin{tikzpicture}[yscale=-1,transform shape]
\pic{orcidlogo};
\end{tikzpicture}
}{|}}}}

\usepackage{hyperref}

\acrodef{LLM}{Large Language Model}
\acrodef{GUI}{Graphical User Interface}
\acrodef{SLM}{Small Language Model}
\acrodef{VPN}{Virtual Private Network}
\acrodef{NLP}{Natural Language Processing}
\acrodef{RNN}{Recurrent Neural Network}
\acrodef{LSTM}{Long Short-Term Memory}
\acrodef{GPU}{Graphics Processing Unit}
\acrodef{CPU}{Central Processing Unit}
\acrodef{ITT}{Inter-Token Time}
\acrodef{LAN}{Local Area Network}
\acrodef{ML}{Machine Learning}
\acrodef{AI}{Artificial Intelligence}
\acrodef{DL}{Deep Learning}
\acrodef{BiLSTM}{Bidirectional Long Short-Term Memory}
\acrodef{IoT}{Internet of Things}
\acrodef{MitM}{Man-in-the-Middle}

\begin{document}

\title{LLMs Have Rhythm: Fingerprinting Large Language Models Using Inter-Token Times and Network Traffic Analysis}

\author{
    \IEEEauthorblockN{Saeif Alhazbi~\orcidicon{0000-0002-7884-5025}\IEEEauthorrefmark{1}, Ahmed Mohamed Hussain~\orcidicon{0000-0003-4732-9543}\IEEEauthorrefmark{2}, Gabriele Oligeri~\orcidicon{0000-0002-9637-0430}\IEEEauthorrefmark{1}, Panos Papadimitratos~\orcidicon{0000-0002-3267-5374}\IEEEauthorrefmark{2}}\\
    \IEEEauthorblockA{
    \IEEEauthorrefmark{1}College of Science and Engineering (CSE), Hamad Bin Khalifa University (HBKU) -- Doha, Qatar
    \\ \{salhazbi, goligeri\}@hbku.edu.qa} \\
    \IEEEauthorblockA{\IEEEauthorrefmark{2}Networked Systems Security Group, KTH Royal Institute of Technology -- Stockholm, Sweden
    \\ahmed.hussain@ieee.org, papadim@kth.se}
}

\maketitle

\begin{abstract}
As Large Language Models (LLMs) become increasingly integrated into many technological ecosystems across various domains and industries, identifying which model is deployed or being interacted with is critical for the security and trustworthiness of the systems. Current verification methods typically rely on analyzing the generated output to determine the source model. However, these techniques are susceptible to adversarial attacks, operate in a post-hoc manner, and may require access to the model weights to inject a verifiable fingerprint. In this paper, we propose a novel passive and non-invasive fingerprinting technique that operates in real-time and remains effective even under encrypted network traffic conditions. Our method leverages the intrinsic autoregressive generation nature of language models, which generate text one token at a time based on all previously generated tokens, creating a unique temporal pattern–like a rhythm or heartbeat–that persists even when the output is streamed over a network. We find that measuring the Inter-Token Times (ITTs)–time intervals between consecutive tokens–can identify different language models with high accuracy. We develop a Deep Learning (DL) pipeline to capture these timing patterns using network traffic analysis and evaluate it on 16 Small Language Models (SLMs) and 10 proprietary LLMs across different deployment scenarios, including local host machine (GPU/CPU), Local Area Network (LAN), Remote Network, and Virtual Private Network (VPN). The experimental results confirm that our proposed technique is effective and maintains high accuracy even when tested in different network conditions. This work opens a new avenue for model identification in real-world scenarios and contributes to more secure and trustworthy language model deployment.
\end{abstract}


\section{Introduction}
\label{sec:intro}

\begin{figure*}[t]
    \centering
    \includegraphics[width=0.8\textwidth]{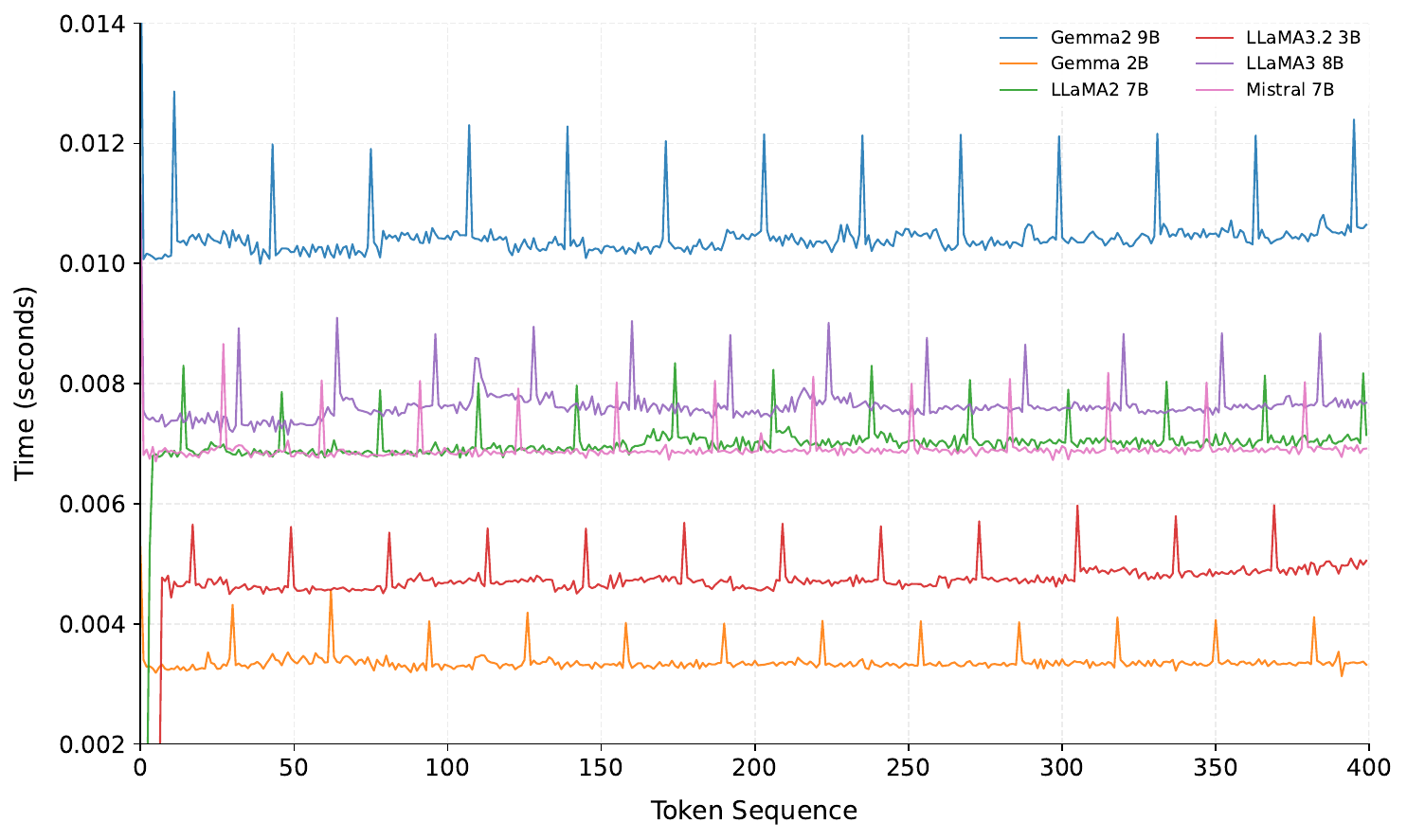}
    \caption{\acp{ITT} of six different \acp{SLM}. Each \ac{SLM} was given the same prompt and instructed to repeat identical text to ensure the generated output was identical across all models. All \acp{SLM} were run on the same \ac{GPU} and machine. Notice the distinct differences in latency, spikes, periodicity, and jitter, which together form each model’s unique timing ``fingerprint''.}
    \label{fig:slm_fingerprint}
\end{figure*}

In recent years, language models, particularly \acfp{LLM}, have experienced accelerated advancement and widespread adoption across various fields. Their remarkable capabilities in language comprehension, reasoning, text, and code generation have unlocked new possibilities for automating and solving complex cognitive tasks~\cite{brown2020language,touvron2023llama}. Due to their computational complexity and resource requirements, these models are typically deployed on expensive specialized hardware in cloud data centers and accessed as a cloud service through web applications or APIs provided by the vendors~\cite{wen2024device}. Building upon these services, third-party platforms integrate multiple state-of-the-art~\acp{LLM} and offer them as a single subscription service to clients. 

However, this remote deployment paradigm introduces critical security and privacy challenges, especially from the client's perspective. A fundamental security concern in this ecosystem is the lack of mechanisms for clients to verify the identity and integrity of the language models they interact with. This becomes critical as individuals and organizations increasingly rely on these models for sensitive tasks, raising concerns about service providers potentially modifying or substituting models without client knowledge or consent. Even in the case of reputable service providers, clients must place considerable trust in them to deliver exactly what they claim to offer in terms of model performance. 

Existing solutions for \acp{LLM} identification broadly fall into two main categories: watermarking and fingerprinting. Watermarking works by embedding imperceptible markers within the model output during the generation process, which can later be detected through algorithmic analysis tools. Yet, this technique is vulnerable to adversarial attacks through paraphrasing or text modification, which can obscure the watermark~\cite{kirchenbauer2023watermark,zhao2024sok}. Unlike watermarking, fingerprinting does not embed markers but rather identifies the model based on the unique inherent characteristics in its generated output. Fingerprinting can be passive, which involves analyzing the model’s output after generation, looking for any statistical or stylistic pattern. However, this is susceptible to text manipulation attacks~\cite{mcgovern2024your}. Active fingerprinting, on the other hand, requires carefully designed prompts to elicit a specific model response for its identification~\cite{pasquini2024llmmap, iourovitski2024hide}. While robust, the active fingerprinting technique is computationally intensive and requires access to the model weights for fine-tuning and response alignment. Overall, both verification techniques operate in a post-hoc manner, analyzing the model output after generation. 

In this paper, we investigate whether autoregressive language models can be uniquely identified from their token generation timing patterns as they stream their responses. Specifically, we measure and analyze \acfp{ITT}---the temporal intervals between consecutive tokens---that arise from the autoregressive generation process to determine if they can serve as a reliable fingerprint. Through extensive analysis, we show that these \acp{ITT} form a unique ``rhythm'' or signature that depends on the model’s architecture, parameter size, and underlying hardware. As illustrated in Fig.~\ref{fig:slm_fingerprint}, \acfp{SLM} with similar parameter sizes or from the same family exhibit a distinct \acp{ITT} pattern different from other models despite generating identical tokens and running on the same \ac{GPU}. Each model has a unique temporal signature, characterized by variations in latency, periodicity, timing spikes, and subtle fluctuations in \acp{ITT} across the generated tokens sequence.  

When autoregressive language models are accessed remotely via encrypted channels, these unique token generation timing patterns are preserved and propagated through the network as the models stream their responses to clients in real-time.  Despite varying network conditions (e.g., latency, jitter, routing) and protocol overhead (e.g., encryption, packetization), these timing characteristics remain observable in network traffic and provide a reliable fingerprint for model identification. 

To capture this fingerprint, we design and implement a \ac{DL}-based pipeline that processes network traffic data and performs feature engineering extracting 36 features. These features are then passed to a hybrid \ac{DL} architecture consisting of \acp{BiLSTM} layers with a multi-head attention mechanism to identify the model. We conduct extensive evaluations of our proposed technique on both open-source \acp{SLM} and proprietary \ac{LLM}. 

Our experiments span a wide range of deployment scenarios, including local GPU/CPU deployment, \ac{LAN}, remote networks, and \ac{VPN}. Across these experiments, the results consistently demonstrate the effectiveness and robustness of our approach in identifying language model families and distinguishing between model variants. These findings provide a new perspective on model identification and ensure greater trust and integrity in their usage. 

Our \emph{contributions} can be summarized as follows:

\begin{itemize}
    \item We demonstrate that autoregressive language models exhibit unique temporal patterns during token generation and propose a novel passive, real-time fingerprinting technique that leverages these unique patterns for model identification in both local and remote network scenarios. 
    \item We design and implement an end-to-end pipeline that processes network traffic, extracts 36 timing and size features to capture the language model's fingerprint and employs a hybrid \ac{BiLSTM}-attention model to classify language models based on these features.
    \item We validate our approach through comprehensive experiments on 16 \acp{SLM} and 10 proprietary \acp{LLM} across various deployment scenarios (local host, \ac{LAN}, remote network, \ac{VPN}). Our results demonstrate the technique’s effectiveness and robustness in identifying both model families and specific variants, even under different network conditions.
\end{itemize}

\textbf{Paper Organization.} Section~\ref{sec:background} provides background and related work; Section~\ref{sec:Preliminaries} formulates the problem; Section~\ref{sec:adver_scenario} presents our scenario and adversary model; Section \ref{sec:methodology} details the proposed methodology and experimental setup; Section~\ref{sec:experimental_result} reports the experimental results and analysis; and finally, Section~\ref{sec:conclusion} concludes the paper.

\section{Background and Related Work} 
\label{sec:background}
The development of language models is built upon decades of research and technological advancement~\cite{zhao2023survey}. In particular, the revolution of language models began with the groundbreaking work in~\cite{bengio2000neural}, which introduced the first probabilistic language model based on neural networks and laid the foundation for using word embeddings and applying \ac{DL} in \ac{NLP}. Following that, the work in~\cite{mikolov2013efficient,pennington2014glove} enhanced word representation to capture semantic relationships more efficiently than earlier techniques. In 2017, the field of \ac{NLP} witnessed a paradigm shift with the introduction of the Transformer architecture~\cite{vaswani2017attention}. Traditional architectures such as \acp{RNN} and \ac{LSTM} suffered primarily from limitations in capturing long-term dependency and had to process text sequentially, which made their training and inference slow and computationally expensive~\cite{hochreiter1997long}. In contrast, the Transformer architecture addressed these issues through self-attention and positional encoding, which enabled parallelization and effective handling of long dependencies. This breakthrough has laid the groundwork for modern \acp{LLM} such as GPT and LLaMA, which scale to billions of parameters. 

\begin{figure*}[t]
    \centering
    \includegraphics[width=0.8\textwidth]{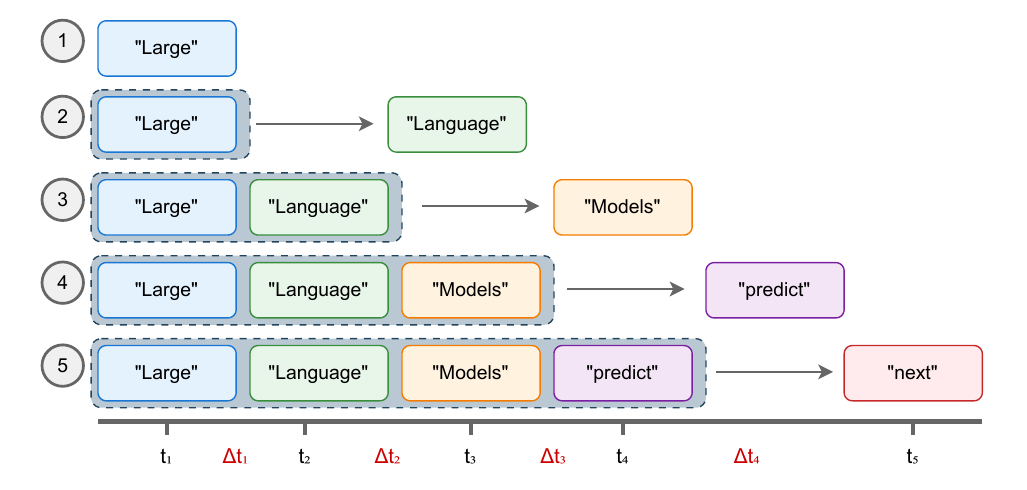}
    \caption{The autoregressive token generation process in \acp{LLM}. At each step (1-5), the model uses all previous tokens (shown in dashed boxes) as a context to predict the next token, then append the new token to the sequence. Generated tokens are shown with corresponding timestamps ($t_1...t_5$). $\Delta t$ represents \acp{ITT} between consecutive tokens generation.}
    \label{fig:autoregressive_process}
\end{figure*}

At their core, \acp{LLM} operate autoregressively, generating text one token at a time based on previously generated tokens and learned contextual embeddings. A token (e.g., a word or subword) is the basic unit of text processed and generated by the language models and is typically selected based on the highest probability of being next in the sequence. This iterative process continues until either a maximum sequence length is reached or a special end-of-sequence token is generated. Fig.~\ref{fig:autoregressive_process} shows the sequential autoregressive token generation process in \acp{LLM}.
At inference time, the generation speed of a language model is primarily determined by factors such as its architecture, parameter size, and hardware parallelization capabilities. Although these large models achieve state-of-the-art performance in a wide range of \ac{NLP} tasks, their computational complexity and massive size require specialized hardware to run efficiently. Hence, most \acp{LLM} are deployed in cloud data centers that can supply the required computational resources.
As the adoption of \acp{LLM} such as ChatGPT, Gemini, and LLaMA becomes widespread, concerns about the security and privacy implications associated with these models have also grown~\cite{yao2024survey, das2024security, yan2024protecting, neel2023privacy}. Issues such as distinguishing between machine- and human-authored content, safeguarding intellectual property, and training models on stolen data are among the primary concerns. Researchers have been actively developing a variety of \acp{LLM} fingerprinting and detection techniques aiming to address these challenges. 

Two primary approaches have emerged in response to these issues: \emph{watermarking} and \emph{fingerprinting}. Watermarking embeds identifying markers into the model-generated output to trace its origin and verify its authenticity. These markers are imperceptible to humans yet detectable through algorithmic methods. For instance, pioneering work by authors in~\cite{kirchenbauer2023watermark} introduced a watermarking framework that modifies the output distribution by selecting a randomized set of "green" tokens before generating each word. Then, during the sampling process, the model softly promotes the use of these green tokens to create a statistical pattern without degrading the text quality. However, such methods are vulnerable to adversarial attacks, including text paraphrasing and token manipulation, which can compromise the watermark's integrity and reliability~\cite{sadasivan2023can}. In contrast, fingerprinting focuses on detecting inherent patterns or characteristics in the model outputs to identify the generative model. Fingerprinting techniques can be categorized broadly into two main approaches: passive and active fingerprinting. 

Passive fingerprinting analyzes the intrinsic characteristics of the model output by examining its lexical and stylistic patterns. For example, authors in~\cite{mcgovern2024your} found that \acp{LLM} produce unique linguistic patterns and writing styles, such that even simple n-gram and part-of-speech distribution analysis can serve as effective fingerprints. These subtle variations in the frequency of specific lexical and syntactic features can differentiate between human- and machine-generated texts, as well as between different model families, such as GPT and LLaMA. The study also found that models within the same family often share similar fingerprints, even across different model sizes. Although this method requires no model modification or special queries, it is vulnerable to adversarial attacks that obscure the fingerprint through text manipulation. Another passive fingerprinting strategy introduced in~\cite{nazari2024llm} exploits memory usage patterns to identify the architectural family of \acp{LLM} deployed on edge and embedded devices. The key idea is that different \ac{LLM} architectures exhibit distinct memory usage patterns during inference. By collecting high-resolution memory usage traces via the \emph{tegrastats} tool and training a \emph{ROCKET} model from the Sktime library, the authors effectively classified even previously unseen \ac{LLM} families with an average accuracy of 92\%. Alternatively, active fingerprinting methods, which require direct interaction with the model, can be further categorized based on their level of access. 

In black-box active fingerprinting, specific queries or prompts are sent to the model to elicit responses that aid in its identification. These methods typically assume limited access to the model--often via an API or by observing its outputs. For example,~\cite{pasquini2024llmmap} proposed LLMmap, a novel black-box active fingerprinting technique that sends carefully crafted queries to the target application and analyzes the responses to identify the specific \ac{LLM} version in use. Their query selection is informed by domain expertise on how \acp{LLM} generate uniquely identifiable responses to thematically varied prompts. The targeted query families include banner grabbing, alignment-based prompts, and malformed queries. With as few as 8 interactions, LLMmap achieves over 95\% accuracy in identifying 42 different \ac{LLM} versions, including both open-source and proprietary models. Similarly,~\cite{iourovitski2024hide} proposed Hide and Seek, a black-box approach using an Auditor-Detective framework. In this method, one \ac{LLM} (the Auditor) generates discriminative prompts, while another \ac{LLM} (the Detective) analyzes the responses to determine family relationships. This approach achieved 72\% accuracy in distinguishing between popular architectures such as LLaMA, Mistral, and Gemma.

On the other hand, white-box active fingerprinting assumes full access to a model's weights and architecture. This level of access enables researchers to deliberately embed a unique signature or pattern into the model during the training or fine-tuning phase for ownership verification. For example, authors in~\cite{russinovich2024hey} introduced novel cryptographic fingerprinting techniques called Chain and Hash to prove the ownership of \ac{LLM} models. Their approach involves generating a set of special questions and answers, then concatenating each question with all questions, all potential answers, and a secret key using a secure hashing function like SHA-256. This fingerprint is then incorporated into the model through fine-tuning, with additional measures like meta-prompts and random padding to enhance the robustness. The verification process is performed by querying the model with the fingerprinting questions and checking whether it produces the expected responses. Similarly, ~\cite{xu2024instructional} proposed a lightweight instruction tuning approach to embed verifiable fingerprints in \acp{LLM}. This technique trains models to generate predetermined outputs when presented with carefully crafted multilingual character sequences (secret keys). Remarkably, using only 60 training instances, this technique demonstrated perfect fingerprint retention across 11 different \acp{LLM}, even after extensive fine-tuning.

Despite these advances, existing \acp{LLM} watermarking and fingerprinting methods have several limitations. Watermarking techniques are often vulnerable to adversarial attacks, while some fingerprinting methods are computationally intensive and require access to model weights. To address these challenges, we propose a novel passive fingerprinting technique that identifies language models in real-time by leveraging their intrinsic \acp{ITT} characteristics, that are inherently difficult to manipulate or obscure via adversarial attacks. Furthermore, our approach is computationally efficient and can be deployed for online detection and monitoring, offering fine-grained fingerprinting capabilities that accurately distinguish between model families and even their variants.

\section{Problem formulation}
\label{sec:Preliminaries}
In this section, we establish the theoretical foundation of the proposed fingerprinting technique. We begin by formalizing the autoregressive generation process inherent in \acp{LLM} and defining the concept of token. We then demonstrate how these tokens are transmitted in network packets, showing how timing patterns and packet sizes can serve as distinctive fingerprints of the underlying model. Finally, we define our classification framework and focal loss objective.

\textbf{Token Definition.} A token is the basic unit of input or output text processed by a language model. Let $\mathcal{M}$ denote a language model with a vocabulary $\mathcal{V}$, such that each token $s_i$ $\in \mathcal{V}$ may represent a word, subword, or individual character, depending on the model's tokenizer. 

\textbf{Token Generation Process.} For each finite output sequence of $N$ tokens generated by $\mathcal{M}$:
\begin{equation*}
    \mathcal{S} = \{ s_1, s_2, \dots, s_N \} \subset \mathcal{V}
\end{equation*}

where each $s_i$ corresponds to the $i$-th token in the generated text. Each token is generated at a specific time $t_i$ so that the overall model output, annotated with timestamps, is given by:

\begin{equation*}
    (s_1,t_1), (s_2,t_2), \ldots, (s_N,t_N)
\end{equation*}

Let $\mathcal{T}$ denote the sequence of timestamps associated with the token generation, defined as:
\begin{equation*}
    \mathcal{T} = \{ t_1, t_2, \dots, t_N \} \in \mathbb{R}^N_+
\end{equation*}

The \acf{ITT} between consecutive tokens is defined as:

\begin{equation*}
    \Delta t_i = t_{i+1} - t_i, \quad i = 1,2,\ldots,N-1.
\end{equation*}

Here, $\Delta t_i$ measures the time required for the model (and its underlying hardware) to generate the next token. In general, in an autoregressive language model $\mathcal{M}$ each token $s_i$ is drawn from the model's vocabulary $\mathcal{V}$ according to:

\begin{equation}
    P(s_i | s_{<i}) = \mathcal{M}(s_1, \dots, s_{i-1})
    \label{eq:token_generation}
\end{equation}

with $s_{<i}$ representing all tokens generated prior to $s_i$. At inference time, the autoregressive language model selects one token at each step using different sampling techniques (i.e., greedy decoding, temperature sampling, or top-k sampling) until an ending criterion is met.

\textbf{From Tokens to Network Packets.} While $\Delta t_i$ captures the generation pattern between tokens on the model side, the output is typically observed via a streaming API or web interface. Hence, in a networking scenario, tokens are bundled into packets for transmission over the network (e.g., LAN or Internet). Therefore, in this practical scenario, the packet is the measurable unit on the client side rather than the token.

Let $\mathcal{P} = \{p_1, p_2, \dots, p_M\}$ denote the sequence of packets received by the client.  Each packet $p_i$ arrives at timestamp $t_{i}^a$ and may contain one or more tokens from $\mathcal{S}$ such that:

\begin{equation*}
    p_i = f_{pack}(s_i, s_{i+1}, ..., s_{i+m})
\end{equation*}

where $f_{pack}$ represents the packetization function and $m$  the number of tokens in packet $i$. Each packet $p_i$ is encrypted before transmission:

\begin{equation*}
    p_i^{enc} = E(p_i, key)
\end{equation*}

with $E$ denoting the encryption function. The packet arrival time $t_i^a$ is different from the generation time due to network latency ($l$) and jitter ($\epsilon$):

\begin{equation*}
    t_i^a = t_i + l + \epsilon
\end{equation*}
On the client side, the inter-arrival time of packets is measured as:

\begin{equation*}
\Delta t_i^p = t_{i+1}^a - t_i^a, \quad i = 1,2,\ldots,M-1
\end{equation*}

This value is affected by both the language model’s \aclp{ITT} ${\Delta t_i}$ and network and protocol overhead (i.e., encryption, packetization, jitter, delays, etc.). Packet sizes are denoted as $\{ |p_i| \}$. 

Despite all these added complexities, we hypothesize that the language model's token generation pattern is preserved and remains detectable from these inter-arrival times ${\Delta t_i^p}$ and packet sizes $\{ |p_i| \}$ unless an obfuscation technique is applied at the server side.

\textbf{Feature Extraction.} We define a feature-extraction function that maps the raw $\{\Delta t_i^p\}$ and $\{ |p_i| \}$ to higher-level features (e.g., burst rate, timing entropy, size-time correlation):

\begin{equation*}
   \phi\bigl(\{\Delta t_i^p\},\,\{|p_i|\}\bigr) \;\longrightarrow\; \mathbb{R}^d
\end{equation*}

In our implementation, we extract 36 engineered features from the raw inter-arrival times 
$\{\Delta t_i^p\}$ and packet sizes $\{|p_i|\}$ (see Appendix~\ref{sec:appendix} for details). For each sliding window $w$ over the network data, we map these raw measurements to a 36-dimensional feature vector $x$ using a feature extraction function $\phi$, i.e.,

\begin{equation*}
    x = \phi\bigl(\{\Delta t_i^p\},\,\{|p_i|\}\bigr)  \in \mathbb{R}^{36}
\end{equation*}

\textbf{Classification Framework.} Given a set of $K$ candidate \acp{LLM}, \{$\mathcal{M}_1,\dots,\mathcal{M}_K$\}, our goal is to learn a classifier

\begin{equation*}
   f: \mathbb{R}^d \,\longrightarrow\, \{1, 2, \ldots, K\}
\end{equation*}.

Here $ \mathbb{R}^d$ denotes the $d$-dimensional feature space derived from the feature-extraction map. We collect a dataset $D$ and label it as:
\begin{equation*}
   \mathcal{D} \;=\; \{(\mathbf{x}_i,\,y_i)\}_{i=1}^{N},
\end{equation*}

with $\mathbf{x}_i\in \mathbb{R}^d$ representing the $i$th sample corresponding to the extracted features vector for a window $w$ of observed network packet sequence (comprising inter-arrival times and sizes), and $y \in \{1,\dots,K\}$ denoting the true label corresponding to the model $\mathcal{M}_k$ that generated that traffic snippet.

Given Dataset $D$, we train a deep neural network model 
$f(\mathbf{x}; \theta)$(parameterized by $\theta$) to output a probability distribution over $K$ classes:
\begin{equation*}
   f(\mathbf{x};\theta)
   \;=\;
   \bigl(
     p_1(\mathbf{x};\theta),\;
     p_2(\mathbf{x};\theta),\;\dots,\;
     p_K(\mathbf{x};\theta)
   \bigr)
\end{equation*}

where $p_k(\mathbf{x};\theta)$ represents the predicted probability that the feature vector  $\mathbf{x}$ originates from  model $\mathcal{M}_k$, and the probabilities satisfy
$\sum_{k=1}^K p_k(\mathbf{x};\theta)=1$.

\textbf{Focal Loss Objective.}  To handle any potential  class imbalance, we use focal loss~\cite{lin2017focal}, which is defined as:

\begin{align}
\label{eq:focal}
   \mathcal{L}_{\text{focal}}(\theta)
   &= - \sum_{i=1}^{N} \sum_{k=1}^{K}
   \alpha_k\, \bigl(1 - p_k(\mathbf{x}_i;\theta)\bigr)^\gamma \mathbf{1}_{\{y_i=k\}} \notag \\
   &\quad\quad\quad\quad\quad\quad\quad\quad\quad\quad\quad\;
   \log\bigl(p_k(\mathbf{x}_i;\theta)\bigr)
\end{align}

where, for each sample $i$:

\begin{itemize}
    \item  $p_k(\mathbf{x}_i;\theta)$ is the predicted probability of class $k$.
    \item  $\alpha_k\in[0,1]$ is an optional weighting factor to mitigate class imbalance.
    \item  $\gamma \ge 0$ is the focusing parameter that increases emphasis on misclassified samples.
    \item $\mathbf{1}_{\{y_i=k\}}$ is the indicator that is $1$ if $y_i = k$ and $0$ otherwise.
\end{itemize}

By minimizing this focal loss, the classifier focuses on hard-to-classify and underrepresented examples while reducing the weight on easily classified examples. 
The final prediction for a feature vector $x$ is given by:
\begin{equation*}
   \hat{y} \;=\; \arg\max_{k \in \{1,\dots,K\}}\, p_k(\mathbf{x};\theta^*)
\end{equation*}

where the optimal model parameters  $\theta^*$ are found by minimizing Equation~\eqref{eq:focal} via gradient-based methods (e.g., Adam or SGD) during training:

\begin{equation*}
   \theta^*
   \;=\;
   \arg\min_{\theta}\;\mathcal{L}_{\text{focal}}(\theta)
\end{equation*}

Once trained, $f(\cdot;\theta^*)$ provides a mapping from the extracted network-traffic features to the predicted \ac{LLM} identity.

\section{Scenario and Adversary Model}
\label{sec:adver_scenario}

\begin{figure*}[t]
    \centering
    \includegraphics[width=0.7\textwidth]{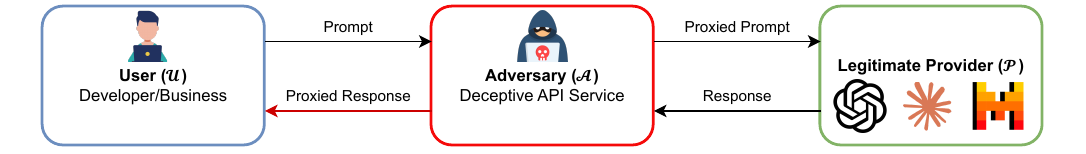}
    \caption{Man-in-the-Middle attack scenario where an active adversary ($\mathcal{A}$) provides a deceptive LLM API service, forwarding the user's prompt requests to a legitimate \ac{LLM} provider ($\mathcal{P}$) and relaying the responses back to the user ($\mathcal{U}$).}
    \label{fig:adversary-model}
\end{figure*}

We consider a scenario involving three entities: the \textit{User} ($\mathcal{U}$), typically a normal user or developer, who interacts with what they believe to be a legitimate \ac{LLM} API service to perform tasks such as text completion or function-calling; the \textit{Legitimate LLM Provider} ($\mathcal{P}$) (e.g., OpenAI, Anthropic, or Mistral), who operates a production \ac{LLM} service through authenticated APIs; and an \textit{Adversary} ($\mathcal{A}$) acting as a Man-in-the-Middle (MitM) in an active role. Fig.~\ref{fig:adversary-model} illustrates the adversarial scenario.

The adversary $\mathcal{A}$ runs a deceptive API service that claims to provide proprietary, state-of-the-art \ac{LLM} functionality, but in reality, $\mathcal{A}$ forwards prompts from $\mathcal{U} $to $\mathcal{P}$ and simply relays the responses back to $\mathcal{U}$. To avoid being unmasked, $\mathcal{A}$ implements a filtering mechanism that detects and blocks any attempts to query model identity before forwarding prompts to $\mathcal{P}$.  In this scenario, since the \ac{LLM} operates as a streaming service, we assume $\mathcal{A}$ is incapable of obfuscating the \ac{LLM} fingerprint (e.g., by randomizing response timing) as such obfuscation would degrade the quality of service, raise suspicion, and compromise the effectiveness of their deception~\cite{serazzi2023impact}. By intercepting the prompts and responses of $\mathcal{U}$, $\mathcal{A}$ not only compromises the security of $\mathcal{U}$, but also infringes on $\mathcal{P}$'s intellectual property rights and terms of service. Furthermore, $\mathcal{A}$ profits by charging users higher fees compared to $\mathcal{P}$'s pricing under the false claim of providing a cutting-edge model. In this context, $\mathcal{U}$ needs to verify the authenticity of the provided \ac{LLM} service before interacting with the API system and sharing sensitive information. Our proposed fingerprinting technique utilizes a \ac{DL} classifier trained on known \ac{LLM} fingerprints to identify the true underlying \ac{LLM}. This approach reveals whether the service is truly a proprietary state-of-the-art model as claimed, or merely proxying requests to an existing commercial \ac{LLM} provider's model.

\section{Methodology}
\label{sec:methodology}
In this section, we discuss in detail our comprehensive experimental framework to study and analyze the unique \ac{ITT} patterns of a variety of \acp{LLM}. We first describe the two types of language models used in our study, along with the different experimental scenarios to evaluate them. Then, we outline the data collection and processing phase, followed by the training and evaluation methodology. Finally, we report on our hardware and software experimental setup. Our investigation involves two primary categories of models: (i) open-source  \acp{SLM} running locally and (ii) proprietary \acp{LLM} accessed via their \ac{GUI} platforms. For each category of language models, we conducted experiments across different scenarios to evaluate the impact of factors such as hardware configurations and network conditions on the models' token generation timing patterns.

\subsection{Open-Source SLMs}
To understand how \acp{LLM} generate their patterns and establish a foundation for our experiment, we collected clean fingerprints from open-source \acp{SLM} deployed locally. Our experiment includes 16 \acp{SLM} spanning five model families: Gemma from Google~\cite{team2024gemma}, Granite3 from IBM~\cite{granite2024granite}, LLaMa from Meta~\cite{touvron2023llama}, Mistral:7B and Ministral:8B from Mistral~\cite{jiang2023mistral}, and Phi from Microsoft~\cite{javaheripi2023phi, abdin2024phi}. These models represent diverse architectures and parameter sizes, ranging from 1B to 9B parameters. Initially, we systematically analyzed \acp{SLM} fingerprints by running identical models locally on both \ac{GPU} and \ac{CPU}. Through this comparative hardware analysis, we aimed to investigate how underlying hardware infrastructure impacts these temporal fingerprints. Then, using GPU deployment, we evaluated the \acp{SLM} across three distinct configurations simulating different real-world scenarios. Starting with \acp{SLM} allowed us to establish a controlled experimental foundation to analyze and trace their temporal fingerprints from source to destination while simultaneously examining the factors impacting these fingerprints. We hypothesize that if we can successfully capture fingerprints generated by these \acp{SLM}, then \acp{LLM} fingerprints would be more distinctive and detectable. This is because \acp{LLM} have significantly more complex architecture and larger parameter sizes (ranging from tens to hundreds of billions) compared to \acp{SLM}, which significantly increases their computational footprint and, as a result, creates more distinct temporal patterns in their token generation process. Our deployment setups include three different scenarios:

\textbf{Local Host Deployment.} In this setup, the client and server are running on the same Linux Ubuntu machine. This configuration eliminates external network overhead since all data transfer occurs within the system, providing an ideal environment for capturing clean \acp{SLM} fingerprints without network variability or latency.
 
\textbf{LAN Deployment.} In this setup, the client and server where \acp{SLM} are deployed operate on two separate machines and are connected via the same \ac{LAN}. On the client side, a Python script automates the process of sending predefined prompts and listening to the streamed response as it is being generated from the Ollama server on the other machine. Meanwhile, \textit{Tshark} captures the incoming packets and filters them according to their IP addresses. The communication between the two devices is not encrypted, and tokens are packetized during transmission. This configuration allows us to observe how the \acp{SLM} fingerprint behaves in a controlled local network environment with minimal latency and variability.
     
\textbf{Remote Deployment via Internet.} In this setup, we aim to investigate the impact of real-world internet conditions on \acp{SLM} fingerprint. To achieve this, we enabled remote access to the Ollama server over the internet using the Cloudflare tunneling service.  The client machine is located in Sweden, while the Ollama server where the \acp{SLM} are hosted and running is located in Qatar. This geographical distribution introduces typical internet-based network conditions to the \acp{SLM} fingerprints.

\textbf{\acp{SLM} Prompting.} We crafted and used a diverse set of prompts, ranging from simple factual questions to complex analytical topics of different lengths, complexity, and response types. Through this methodology, we aimed to simulate realistic usage of these \acp{SLM} and ensure that each model demonstrates its unique token generation behavior under different prompting conditions.
 
\subsection{Proprietary LLMs}
Next, we extended our experiment to include real-world scenarios by accessing proprietary \acp{LLM} over the internet. We selected three families of widely-used proprietary models: OpenAI's GPT series (including GPT-4, GPT-4o, and GPT-4o mini), Anthropic's Claude (Sonnet 3.5, Opus, and Haiku), and Mistral (Pixtral Large, Mistral Large2, Mistral Small, and Mistral Nemo). At the time of writing, these models represent state-of-the-art \acp{LLM}, featuring diverse architectures, parameter sizes, and operational infrastructures. We captured the network traffic of these models and trained a hybrid \ac{DL} model, then tested the same trained model under the following different scenarios: 

\textbf{Different Day.} To evaluate the consistency and persistence of the \acp{LLM} fingerprints over time, we collected the training data on one day and the test data on a different day. Through this experimental design, we aim to examine whether \acp{LLM} fingerprints remain stable despite temporal variations in the models' operational environments, including variability in server load and network conditions at both the client and server endpoints.

\textbf{Different Network.} Network conditions can significantly affect the quality of observed \acp{LLM} fingerprints. To evaluate the impact of network conditions and to validate the network-agnostic nature of \acp{LLM} fingerprints, we trained our detection model using data collected from one network location and tested it using data collected from a different geographical location. This cross-network evaluation assesses the model's ability to generalize across different network environments. In particular, we investigate whether the fingerprint characteristics remain distinguishable despite inherent variations in network conditions such as latency, jitter, and bandwidth between different locations.

\textbf{VPN.} To further validate our proposed technique, we conducted an additional experiment under more complex network conditions. Specifically, we accessed the proprietary \acp{LLM} using \ac{VPN} with an exit node located in a different geographical region. By introducing this proxy layer, we aimed to determine whether the \acp{LLM} fingerprint remains observable despite the additional encryption, routing, and processing overhead introduced by \ac{VPN}. In essence, we evaluated how well the detection model trained under normal network traffic conditions performs in this more challenging scenario.

\textbf{\acp{LLM} Prompting.} Similar to open-source\acp{SLM} we crafted diverse prompts to interact with the proprietary \acp{LLM} through their \ac{GUI} website. These prompts varied in length, complexity, and topic, ranging from simple question-answer exchanges to extended chains of prompts that establish deep contextual dependencies. With this comprehensive prompting technique, we aimed to capture \acp{LLM} fingerprints under realistic usage scenarios that reflect typical user interaction with such models. 

The primary goal of performing these experimental scenarios is to demonstrate that \ac{LLM} fingerprints are inherent characteristics of the models themselves, persisting regardless of temporal variations, network configurations, and routing infrastructures.

\subsection{Data Collection and Processing}
In the data collection phase, we acquired data using a \emph{wired} connection throughout the different experiments, as it provides greater stability and reliability in measuring traffic patterns. We applied filters in \textit{Tshark} and \textit{Wireshark} to capture only inbound, data-only traffic—excluding server-related and control packets—to focus solely on the data packet patterns relevant to the analysis. The captured network traffic raw data was stored in a PCAP file and consisted of traffic packet sizes and their arrival times. As described in Algorithm~\ref{alg:DataPreprocessing}, the data processing pipeline begins with the extraction and pre-processing of raw network traffic data. Specifically, we extract two fundamental components from the raw network traffic: packet arrival timestamps ($\mathcal{T}_i$) and corresponding packet sizes $\mathcal{P}_i$. From the timestamps, we compute the inter-arrival times ($\Delta \mathcal{T}_i$) between consecutive packets to capture the temporal signature of the \acp{LLM} and form the foundation for our fingerprinting methodology's feature extraction process. Following this phase, we perform data de-noising to remove anomalies and address any missing or invalid values. This preprocessing step ensures that our subsequent feature extraction phase is based on clean and reliable data that accurately represent each model's generation pattern.

\begin{algorithm}[!h]
\caption{Data Extraction and Preprocessing}
\label{alg:DataPreprocessing}
\begin{algorithmic}[1]
\STATE \textbf{Input:}
\STATE \hspace{1em} $N$: Number of LLMs
\STATE \hspace{1em} $\mathcal{D}_i$: Raw network traffic data for each LLM $L_i$, where $i = 1, 2, \dots, N$
\STATE \textbf{Output:}
\STATE \hspace{1em} Cleaned inter-arrival times $\Delta \mathcal{T}_i$ and packet sizes $\mathcal{P}_i$ for each LLM
\vspace{0.5em}
\FOR{each LLM $L_i$, $i = 1$ to $N$}
    \STATE Extract packet arrival times $\mathcal{T}_i = \{ t_1, t_2, \dots, t_n \}$ from $\mathcal{D}_i$.
    \STATE Extract packet sizes $\mathcal{P}_i = \{ p_1, p_2, \dots, p_n \}$ from $\mathcal{D}_i$.
    \STATE Compute inter-arrival times $\Delta \mathcal{T}_i = \{ \Delta t_1, \Delta t_2, \dots, \Delta t_{n-1} \}$ where $\Delta t_i = t_{i+1} - t_i$.
    \STATE Clean and preprocess data:
    \STATE \hspace{1em} -- Remove anomalies and outliers.
    \STATE \hspace{1em} -- Handle missing or invalid values.
\ENDFOR
\end{algorithmic}
\end{algorithm}

\subsection{Features Engineering}
Next, we apply feature engineering to the extracted raw network traffic data. As the \acp{LLM} responses are transmitted over networks, they experience various noise factors and delays, making it infeasible to rely only on raw data (i.e., packet inter-arrival times and packet sizes) for accurate model identification. To address this limitation, we implement a feature engineering process as illustrated in Algorithm~\ref{alg:FeatureExtraction}. The primary objective of feature engineering is to enhance data representation, reduce noise, and better capture the underlying patterns in the stream of \acp{LLM} packets. Using an iterative empirical process based on the two essential features--inter-arrival times and packet sizes--we extract a total of 36 engineered features that focus on metrics revealing the model's token generation behavior. These features span six main categories: rate and throughput metrics (e.g., maximum burst rate and packet rate); inter-arrival time statistics (e.g., mean inter-arrival time and percentile distributions); pattern and regularity metrics (e.g., timing regularity and permutation entropy); timing change dynamics (e.g., mean time change and timing acceleration); correlation and combined metrics (e.g., size-time correlation and size-time products); and burstiness and entropy metrics (e.g., burstiness measure and inter-arrival time entropy). All these features collectively measure different aspects of the model's token generation behavior. Our aim is to recognize the ``rhythm'' of each \ac{LLM} by detecting and analyzing the temporal changes in network activity within a small time window.
We apply a sliding window of 0.5 seconds with a step size of 0.1 seconds to the raw network traffic data, which consists of packet sizes and inter-arrival times. These parameters were determined through extensive empirical evaluation, during which we tested various combinations. Within each data window, we compute a set of statistical and temporal features for that sample of data and label the resulting feature vector with the corresponding \acp{LLM} class $L_i$. The complete mathematical formulas and detailed calculations for all derived features are provided in Appendix~\ref{sec:appendix}.

\begin{algorithm}[!h]
\caption{Feature Engineering}
\label{alg:FeatureExtraction}
\begin{algorithmic}[1]
\STATE \textbf{Input:}
\STATE \hspace{1em} Cleaned inter-arrival times $\Delta \mathcal{T}_i$ and packet sizes $\mathcal{P}_i$ for each LLM
\STATE \hspace{1em} $w$: Sliding window size
\STATE \hspace{1em} $s$: Step size for sliding window
\STATE \textbf{Output:}
\STATE \hspace{1em} Feature vectors $\mathbf{x}_i$ labeled with LLM $L_i$
\vspace{0.5em}
\FOR{each LLM $L_i$, $i = 1$ to $N$}
    \FOR{each window $w$ over $\Delta \mathcal{T}_i$ and $\mathcal{P}_i$ with size $w$ and step $s$}
        \STATE Extract features:
        \STATE \hspace{1em} -- Statistical features (mean, variance, percentiles) of $\Delta t$ and $p$.
        \STATE \hspace{1em} -- Temporal features (burstiness, entropy).
        \STATE \hspace{1em} -- Advanced features (permutation entropy, LIS).
        \STATE Label feature vector $\mathbf{x}_i$ with LLM $L_i$.
    \ENDFOR
\ENDFOR
\end{algorithmic}
\end{algorithm}

\subsection{Training}
\begin{figure*}[t]
    \centering
    \includegraphics[width=0.9\textwidth]{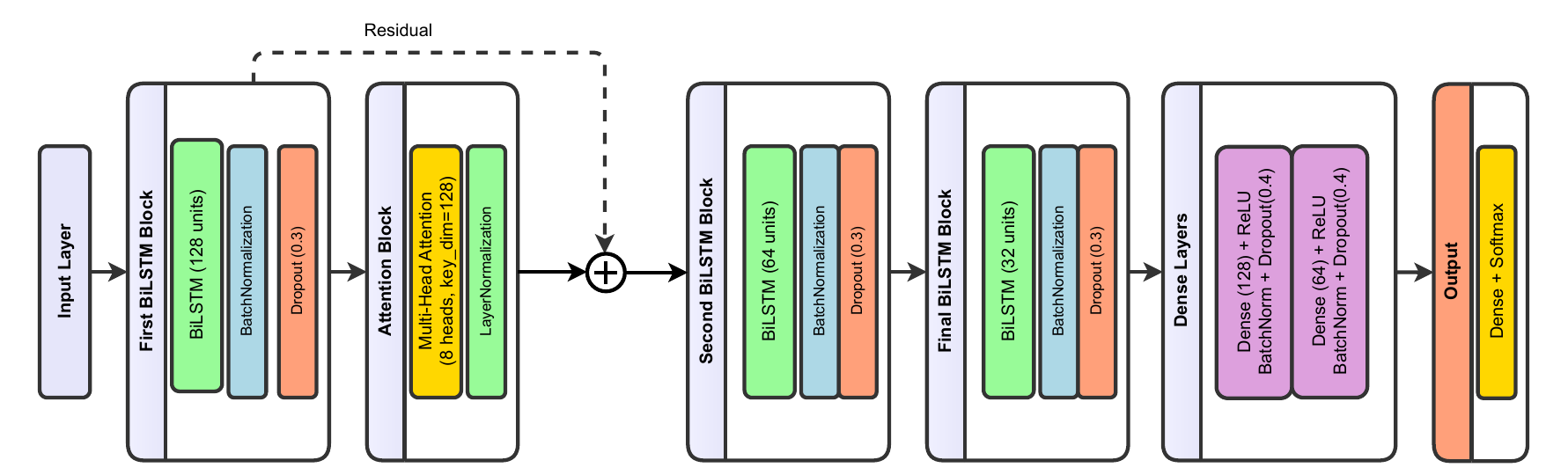}
    \caption{Attention-based \ac{BiLSTM} architecture for \acp{SLM} and \acp{LLM} traffic classification.}
    \label{fig:classification_model}
\end{figure*}

We designed a hybrid \ac{DL} architecture that combines multiple \ac{BiLSTM} blocks with a multi-head attention mechanism to capture sequential long-term dependencies in network traffic patterns while focusing on the most discriminative features. This hybrid architectural approach has been successfully adopted in several contributions to network traffic analysis~\cite{wang2025bcba,yao2019identification}.  As illustrated in Fig.~\ref{fig:classification_model} our model consists of three \ac{BiLSTM} blocks with decreasing units to progressively refine the temporal features throughout the network. After each block, we added batch normalization to stabilize training and dropout layers (rate = 0.3) to prevent overfitting. Following the first \ac{BiLSTM} block, an 8-head attention mechanism with a key dimension of 128 is applied, followed by batch normalization layer. A residual connection then adds the attention output back to the initial \ac{BiLSTM} block's output to preserve the original temporal features and ensure efficient gradient flow before passing them to the second \ac{BiLSTM} block. Finally, the architecture concludes with two dense layers (128 and 64 units) followed by batch normalization and a softmax layer to produce the final classification probabilities. 

The training pipeline, outlined in Algorithm~\ref{alg:ModelTraining} consists of several stages optimized to handle network traffic data. The process starts with data preparation, where the input dataset $\mathcal{D}$ containing the feature vector $\mathbf{x}_i$ and corresponding labels $\mathbf{L}_i$ is partitioned into training and validation sets. Next, the data is preprocessed using a per-sample normalization technique to standardize each input within a time window independently, preserving the relative temporal patterns while mitigating the effects of varying network conditions. To address the class imbalance problem where some \acp{LLM} have more data samples than others, we used two strategies: class weighting and focal loss. This combination prevents bias towards overrepresented classes and maintains sensitivity to undersampled events. During data preparation, we again use the sliding window technique with window size $w = 128$ and step size $s = 4$ to segment the temporal sequences. This creates overlapping windows that preserve the temporal continuity of the \ac{LLM} generation patterns and provide sufficient training samples. We selected a batch size of 64, and trained the model for 30 epochs, as the standard deviation of accuracy in the last 10 epochs remained below 0.01, indicating convergence and training stability. It is worth noting that upon completion of the training phase, we tested the model on a newly collected dataset for each testing scenario.

\begin{algorithm}[!h]
\caption{Model Training}
\label{alg:ModelTraining}
\begin{algorithmic}[1]
\REQUIRE
    \STATE Dataset $\mathcal{D} = \{ (\mathbf{x}_i, L_i) \}$ of feature vectors and labels
    \STATE Number of training epochs $T$
    \STATE Hyperparameters (learning rate $\eta$, batch size, etc.)
\ENSURE
    \STATE Trained classification model $f$
\vspace{0.5em}
\STATE Split $\mathcal{D}$ into training set $\mathcal{D}_{\text{train}}$ and validation set $\mathcal{D}_{\text{val}}$
\STATE Preprocess data with per-sample normalization and adaptive noise

\textcolor{gray}{$\triangleright$ \ Compute class weights for imbalanced data}
\STATE Calculate class frequencies from training labels
\STATE Set $w_i = \sqrt{\max(\text{freq}) / \text{freq}_i}$ for each class $i$

\textcolor{gray}{$\triangleright$ \ Model initialization and compilation}
\STATE Initialize model $f$ with BiLSTM-Attention architecture
\STATE Initialize Focal Loss with $\alpha=0.25$, $\gamma=2.0$, and class weights $w_i$
\STATE Set Adam optimizer with learning rate $\eta$
\FOR{epoch $t = 1$ to $T$}
    \STATE Train model $f$ on $\mathcal{D}_{\text{train}}$
    \STATE Validate model $f$ on $\mathcal{D}_{\text{val}}$
    \STATE Update model parameters based on loss
\ENDFOR
\end{algorithmic}
\end{algorithm}

\subsection{Evaluation}
Given the imbalanced nature of the dataset after applying the preprocessing phase, using traditional accuracy metric can be misleading to report on the model's performance. Instead, we consider the following  evaluation metrics to assess the model's classification performance:

\begin{equation}
    \text{Precision} = \frac{\text{TP}}{\text{TP} + \text{FP}}
\end{equation}

where precision measures the proportion of correct positive predictions, reflecting the model's accuracy in identifying specific \ac{LLM} $x$ such that out of all times the model identified $x$ how often it was correct.

\begin{equation}
    \text{Recall} = \frac{\text{TP}}{\text{TP} + \text{FN}}
\end{equation}

where recall measures the model's ability to find all instances of a specific LLM $x$, such that out of all actual occurrences of $x$ in the data, how many were correctly identified by the model.

\begin{equation}
    \text{F1 Score} = \frac{2 \times \text{Precision} \times \text{Recall}}{\text{Precision} + \text{Recall}}
\end{equation}

where the F1 score provides a balanced metric for the harmonic mean of precision and recall, especially in cases when evaluating model performance across different \acp{LLM} with varying frequencies in the dataset. In addition to the previous metrics, we present a confusion matrix for each conducted experiment to provide a comprehensive evaluation of the model's classification performance across all classes. The evaluation process, as shown in Section~\ref{sec:experimental_result}, is conducted on completely new collected data.

\subsection{Experimental Setup}
\label{sec:exp-setup}
We evaluated our proposed solutions using both open-source \acp{SLM} and proprietary \acp{LLM}. Specifically, we deployed 16   \acp{SLM} running locally on consumer-grade GPUs and CPUs. These open-source lightweight models were installed using Ollama (version 0.3.14), an open-source platform that simplifies the installation and running of \acp{LLM} locally. In contrast, the proprietary models, including ChatGPT, Mistral, and Claude, were accessed through their web-based \ac{GUI}.

\subsubsection{Hardware Configuration}
For data collection, processing, deployment of open-source \acp{SLM}, and training of the classification model, we used a local machine with the following specifications. Our system was equipped with an NVIDIA RTX 4090 GPU, providing 24 GB of VRAM to support high-speed inference and efficient handling of \acp{SLM} parameters and architectures. This GPU was used for both \ac{SLM} deployment and training of the classification models at a later stage. The system's processor was an Intel  Core i9-13900K featuring 24 cores (8 performance cores and 16 efficiency cores) with a maximum clock speed of 5.8 GHz. To meet the memory demands for our computational task, the machine was equipped with 128 GB of DDR5 RAM operating at 5200 MHz. 

\subsubsection{Software Setup}
The host machine was running Ubuntu 24.04.1 LTS. Automation of tasks such as prompting the \acp{SLM}, data collection, and processing was implemented using scripts written in Python 3.11.5. Several Python libraries were used, including \emph{requests} for HTTP communication with the Ollama server running on port 11434 (configured to serve all 16 models), \emph{json} for parsing model responses, \emph{csv} for data logging and storage, and \emph{datetime} for precise timestamp handling. To capture network traffic, we integrated \emph{tshark} directly within the Python script for the locally deployed \acp{SLM}, while \emph{Wireshark} was used for proprietary models during the prompting process \cite{wireshark}. For implementing and training the \ac{DL} model, we used \emph{TensorFlow 2.15.0} managed through \emph{Miniconda} environment manager (version 24.1.2). As for remote access to the Ollama server and to prompt the locally deployed open-source \acp{SLM} over the internet, we used the Cloudflare Tunneling service to establish a secure connection to the local machine. During VPN-based testing, we used Surfshark VPN to introduce network obfuscation and evaluate VPN network impact on the model classification performance.

\section{Experimental Results} 
\label{sec:experimental_result}
In this section, we present our experimental results by evaluating our proposed fingerprinting technique on two categories of language models: open-source \acp{SLM} running locally, and proprietary \acp{LLM} accessed through their websites. For \acp{SLM}, we first investigated their temporal patterns across different deployment scenarios, including local hardware configurations (GPU/CPU), \ac{LAN} environment, and remote network access, where we developed a \ac{DL} classification model to establish the feasibility of fingerprinting in challenging network conditions. For proprietary \acp{LLM}, we trained a classification \ac{DL} model and evaluated its robustness by testing it on data collected on a different day, in a different network, and through \ac{VPN} connection.

\subsection{Open-Source SLMs}
\label{sec:experimental_result_slm}
As an experimental baseline to understand and measure the unique \acp{ITT} that serve as a fingerprint for generative language models, we conducted experiments on 16 open-source \acp{SLM} from five leading companies in language model development. Due to hardware constraints and the high memory and \ac{GPU} requirements of larger open-source \ac{LLM} (e.g., Gemma 27B~\cite{team2024gemma}, or LLaMA 90), we experimented with model sizes ranging from 1B to 9B parameters. These models are considered small and are computationally feasible to run on consumer-grade hardware. We studied these \acp{SLM} across three different deployment scenarios: (i) local host where both client and server run on the same machine, (ii) \ac{LAN} where both run on separate machines within the same network and (iii) remote deployment where client and server communicate over the internet. By adopting this controlled incremental approach, the objective is to establish a ground truth for our investigation and study the phenomenon analytically and progressively before applying any complex machine-learning techniques.

\textbf{Local Host Deployment.} In this scenario, both the server where the \acp{SLM} inference is performed, and the client prompting the models reside in the same machine. This setup allowed us to collect a clean fingerprint by eliminating any network-related factors and focusing purely on the intrinsic \acp{ITT} specific to each language model. To study the hardware impact on model generation characteristics, we ran the same experiment on both \ac{GPU} and \ac{CPU} installed in the same machine. By running \acp{SLM} on the \ac{CPU}, we considered the worst-case scenario where such models may be deployed on hardware with significant resource constraints. All \acp{SLM} were prompted with identical prompts to ensure fairness, consistency, and validity in the comparison across all models. During the response generation process, the Ollama server recorded each token with \emph{created\_at} field indicating the creation timestamps. We used this feature to compute the overall mean \acp{ITT} for each \ac{SLM} across all generated outputs.

\subsubsection*{Deployment on GPU}
\begin{figure}[t]
    \centering
    \includegraphics[width=0.5\textwidth]{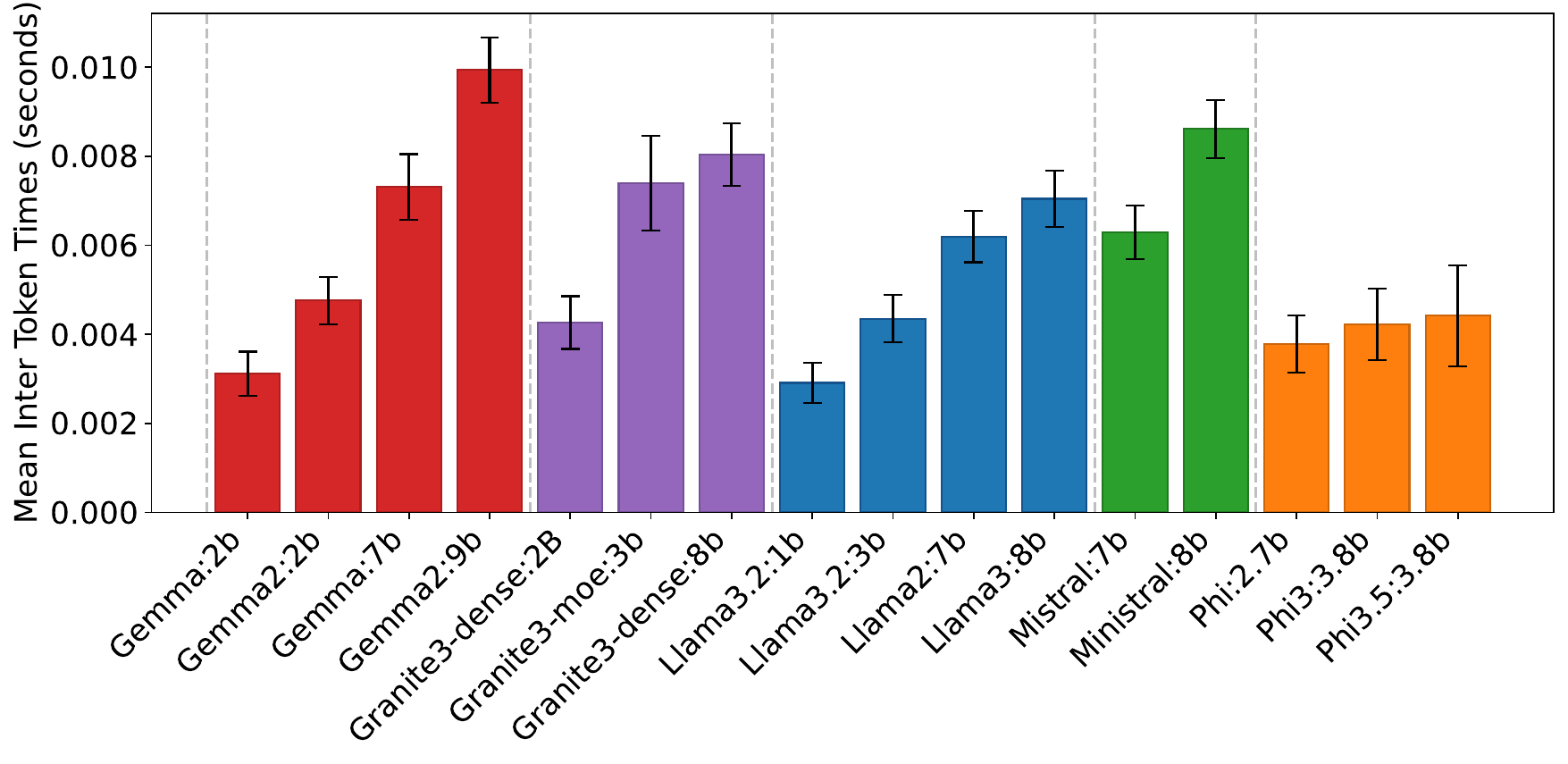}
    \caption{Mean \acp{ITT} for different \acp{SLM} families deployed on \ac{GPU}. Colors represent the model family. Error bars represent the standard deviation from the mean.}
    \label{fig:gpu-timing}
\end{figure}

\begin{figure}[t]
    \centering
      \includegraphics[width=\linewidth]{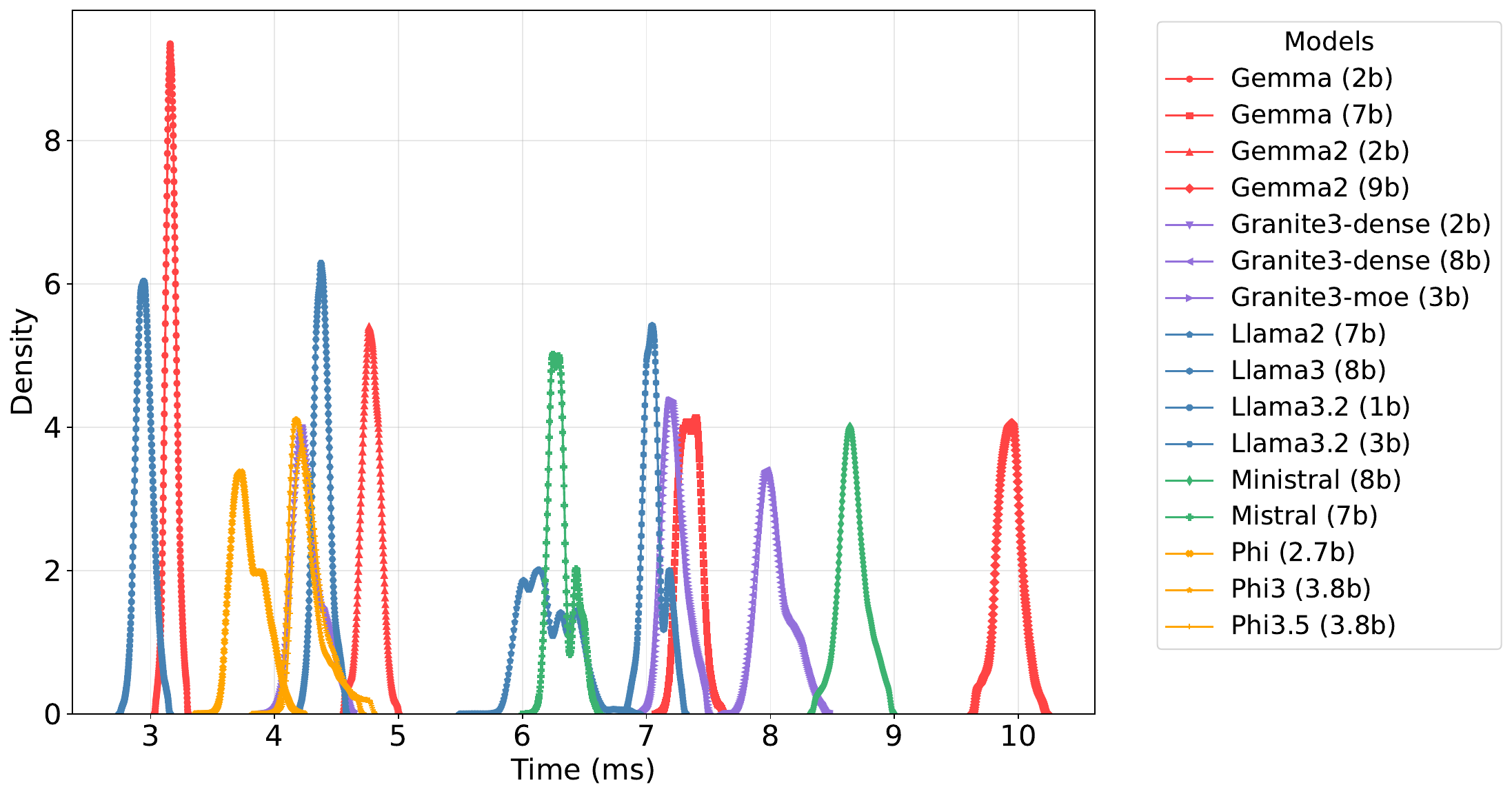}
    \caption{Probability density function of \acp{ITT}
 under \acp{SLM} local host deployment.}
    \label{fig:model_performance_pdf_single}
\end{figure}

First, we analyzed \acp{ITT} for the 16 \acp{SLM} deployed on the \emph{NVIDIA RTX 4090} \ac{GPU}. As shown in Fig.~\ref{fig:gpu-timing} each model exhibits a distinct \ac{ITT} profile different from the other models, even those within the same family demonstrate a distinguishable temporal signature. Furthermore, we observed a clear correlation between model size and mean \ac{ITT} such that models with larger parameter sizes tend to have longer mean \ac{ITT} as they require more  computational resources. This trend is particularly prominent within models of the same family and holds true across all families, though with varying degrees of scaling proportional to model size. For example, the Gemma family which has the widest temporal range, demonstrate the most pronounced size-dependent linear scaling relationship with Gemma2 (9B), showing the longest mean \ac{ITT} of approximately 0.0099 seconds nearly three times that of its smaller (2B) parameter variants Gemma1 (2B), which averages around 0.0031 seconds. Similarly, the LLaMA models demonstrate scaling with a moderate slope from 0.0029 seconds for LLaMA3.2 (1B) to 0.0070 seconds for LLaMA3 (8B). On the other hand, the Phi model's family shows a remarkable efficiency consistency across their variants with \acp{ITT} differences (Phi1 (2.7B): 0.0038s, Phi3 (3.8B): 0.0042s, Phi3.5 (3.8B): 0.0044s). 

The standard deviations for \ac{GPU} deployments are notably low, indicating stable and consistent \acp{ITT}. To complement our analysis and obtain a deeper understanding of the temporal patterns, we plotted the distribution of \acp{ITT} as probability density functions as illustrated in  Fig.~\ref{fig:model_performance_pdf_single}. As confirmed by the previous figure's result, each model exhibits a unique temporal distribution pattern with minimal overlap, even among models of similar parameter sizes from different families. The shapes of these distributions reveal another unique characteristic such that some models like LLaMA3.2 (1B) and Gemma (2B) show sharp, narrow peaks indicating very consistent timing, while others like LLaMA2 (7B) show broader distributions indicating greater timing variability. These variations in distribution shapes are observable both across models of similar parameter sizes and among variants within the same model family. This result reveals several key insights about model generation patterns. 

While architectural similarities within a model family can create shared temporal behavior patterns, each model still produces its own unique, identifiable signature due to differences in parameter sizes and specific optimization techniques. Furthermore, the unique distribution shapes observed between models of similar parameter sizes suggest that these differences stem fundamentally from the architectural design of these models. Overall, these findings establish a strong foundation for our hypothesis that autoregressive language models can be identified through a novel lens using their temporal pattern during the token generation process. This approach enables discrimination not only between model families but also among specific variants within them, potentially serving as a complementary method to current fingerprinting techniques.

\subsubsection{Deployment on CPU}
\begin{figure}[t]
    \centering
    \includegraphics[width=0.5\textwidth]{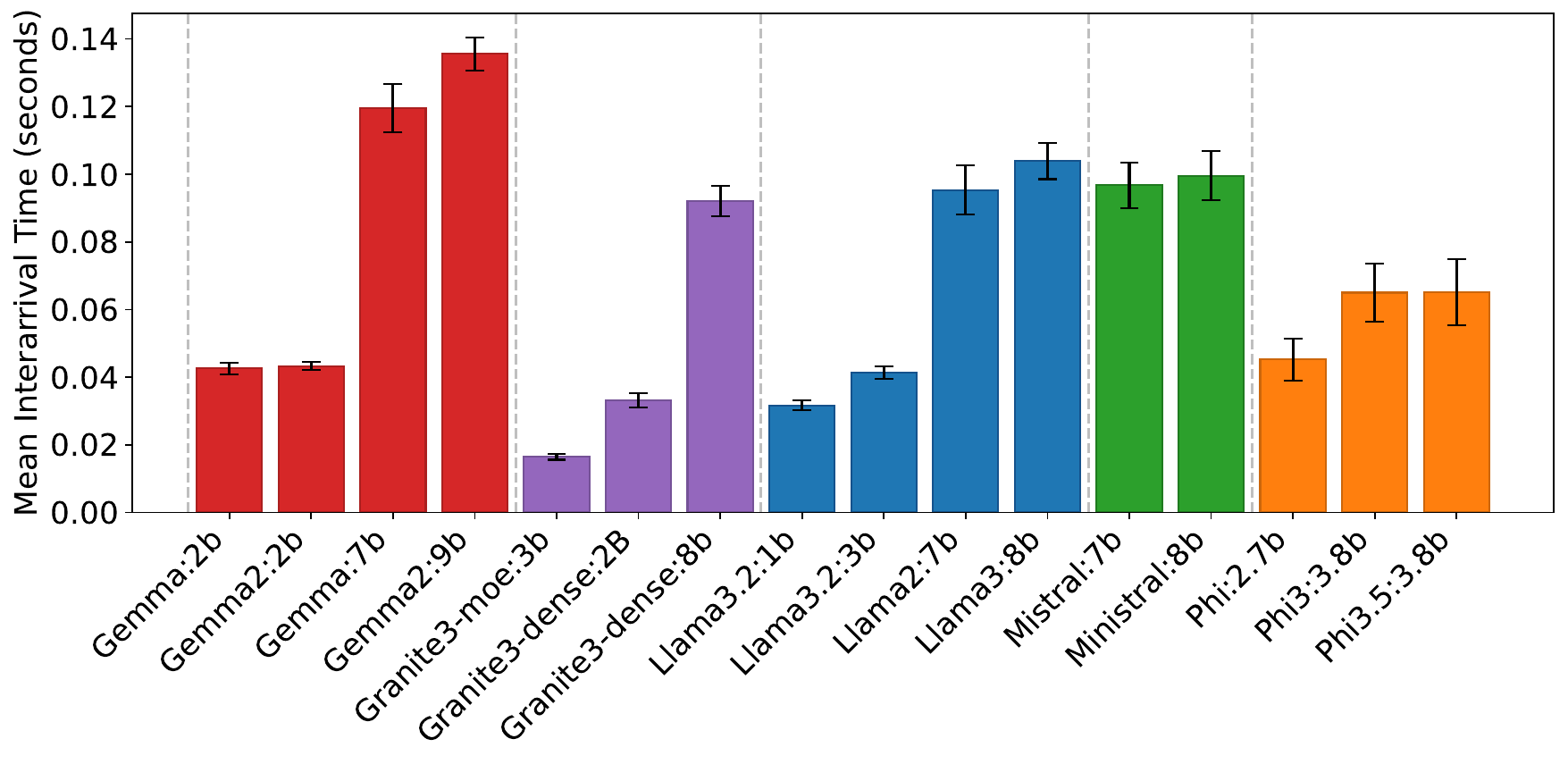}
    \caption{Mean \acp{ITT} for different \acp{SLM} families deployed on \ac{CPU}. Colors represent the model family. Error bars represent the standard deviation from the mean.}
    \label{fig:cpu-timing}
\end{figure}

As anticipated, due to the limited parallel processing capabilities of the \ac{CPU}, the mean \acp{ITT} for \acp{SLM} inference are significantly longer compared to \ac{GPU} execution times, as shown in  Fig.~\ref{fig:cpu-timing}. Specifically, the mean \acp{ITT} on \ac{CPU} range from 0.02 to 0.14 seconds, which is approximately an order of magnitude slower than \ac{GPU} execution times of 0.003 to 0.01 seconds. Furthermore, the relation between model parameter size and inference speed is amplified on \ac{CPU}, with larger models experiencing proportionally greater slowdowns compared to their performance on \ac{GPU}. Moreover, \ac{CPU} performance shows more timing variability demonstrated by larger error bars in the measurement. This trend is more noticeable in larger model families like Gemma and LLaMA. These results demonstrate that hardware configuration is a fundamental significant factor in determining \acp{ITT} of any language model alongside other characteristics such as model architecture, number of parameters, and the optimization techniques used.  However, this hardware dependency does not prevent the practical applicability of our approach, as fingerprinting can be calibrated to the deployed hardware, which typically remains unchanged in the production environment, and can be recalibrated if hardware changes occur.

\textbf{Network Impact.} Most language models, particularly those with large weights, are accessed remotely over the internet as cloud-based services since they require specialized high-end GPU and computing infrastructure, which is typically not available to end-users. Therefore, we aim to investigate how protocol overhead and network conditions may affect the differentiability of language models based on their generation timing patterns when their responses are transmitted to clients. We test \acp{SLM} in two networking scenarios where language models could be deployed: (i) a controlled \ac{LAN} where both the server running the model and the client reside within the same network, and (ii) a remote deployment environment where the models are hosted on an external network and accessed through the internet.

\subsubsection*{LAN Deployment}
\begin{figure}[!htbp]
    \centering
    \includegraphics[width=0.5\textwidth]{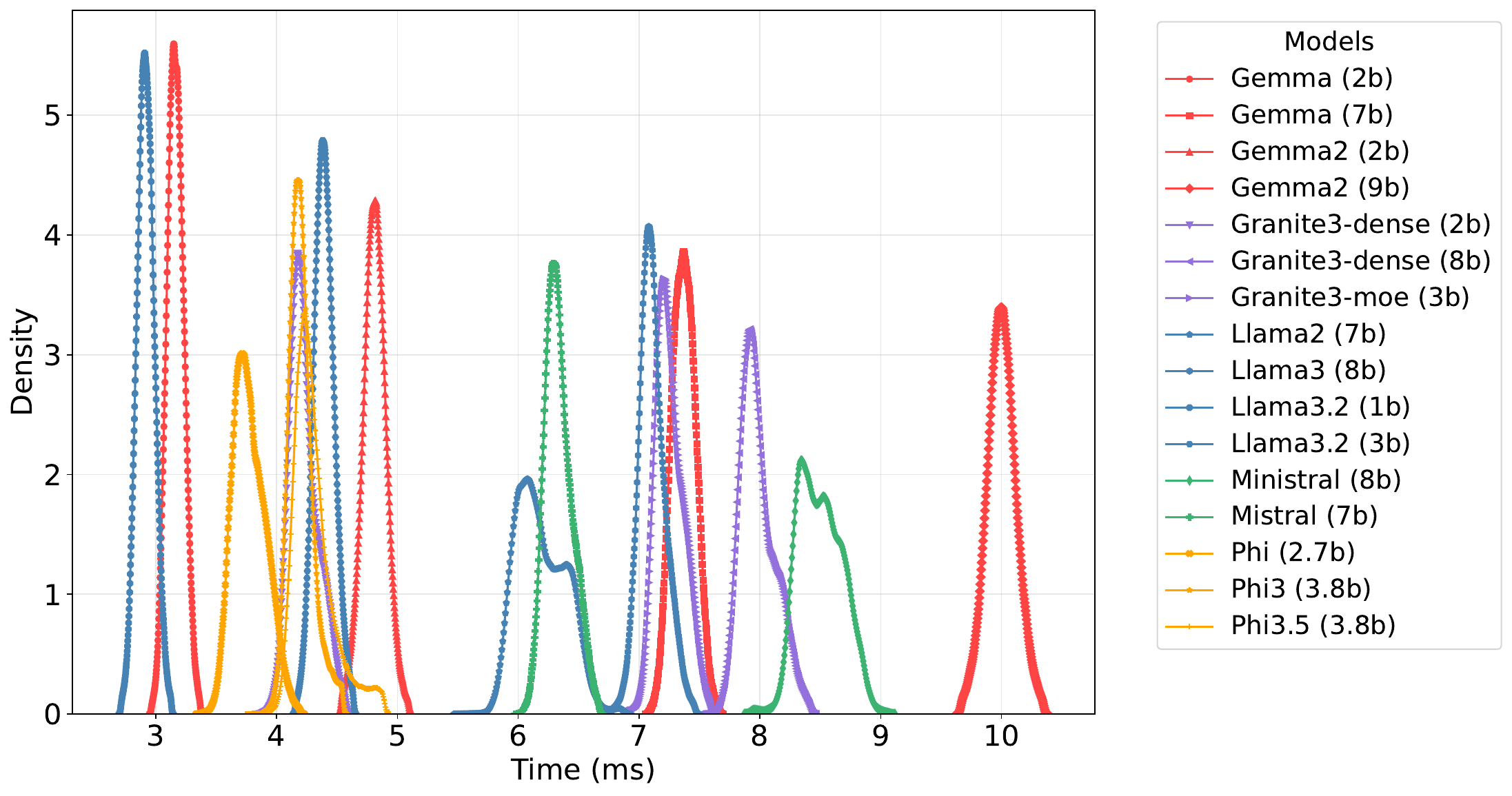}
    \caption{Probability density function of \acp{SLM} packet inter-arrival times under \ac{LAN} deployment conditions.}
    \label{fig:pdf_lan}
\end{figure}

As we move beyond local host inference, network conditions and protocol overhead introduce additional complexity. We progressively evaluate how this may impact \acp{ITT} by testing that within a controlled \ac{LAN} environment with minimal latency and variability. Such an environment closely resembles edge computing and \ac{IoT} networks, where devices communicate locally. In this scenario, two machines communicate directly within the same subset network without any encryption, with one acting as a client and the other one as a server. Tokens generated by the Ollama server are packetized before transmission, simulating a real-time streaming scenario. In this context, a packet might contain one or more tokens generated by the \ac{SLM}; as a result, on the client side, we instead shift our analysis to measure the inter-arrival times between consecutive packets. Fig.~\ref{fig:pdf_lan} displays the probability distribution function of the inter-arrival time of packets for each \ac{SLM}. Compared to the previous result obtained within the local host machine (\ac{GPU} based scenario) in Fig.~\ref{fig:gpu-timing}, the LAN network environment introduces minimal additional variability, resulting in a slight shift and overlaps in the timing distributions. However, the general model groupings and relative performance characteristics remain recognizable despite these network effects. 

\subsubsection*{Remote Network (Internet)}
\begin{figure}[!htbp]
    \centering
    \includegraphics[width=0.5\textwidth]{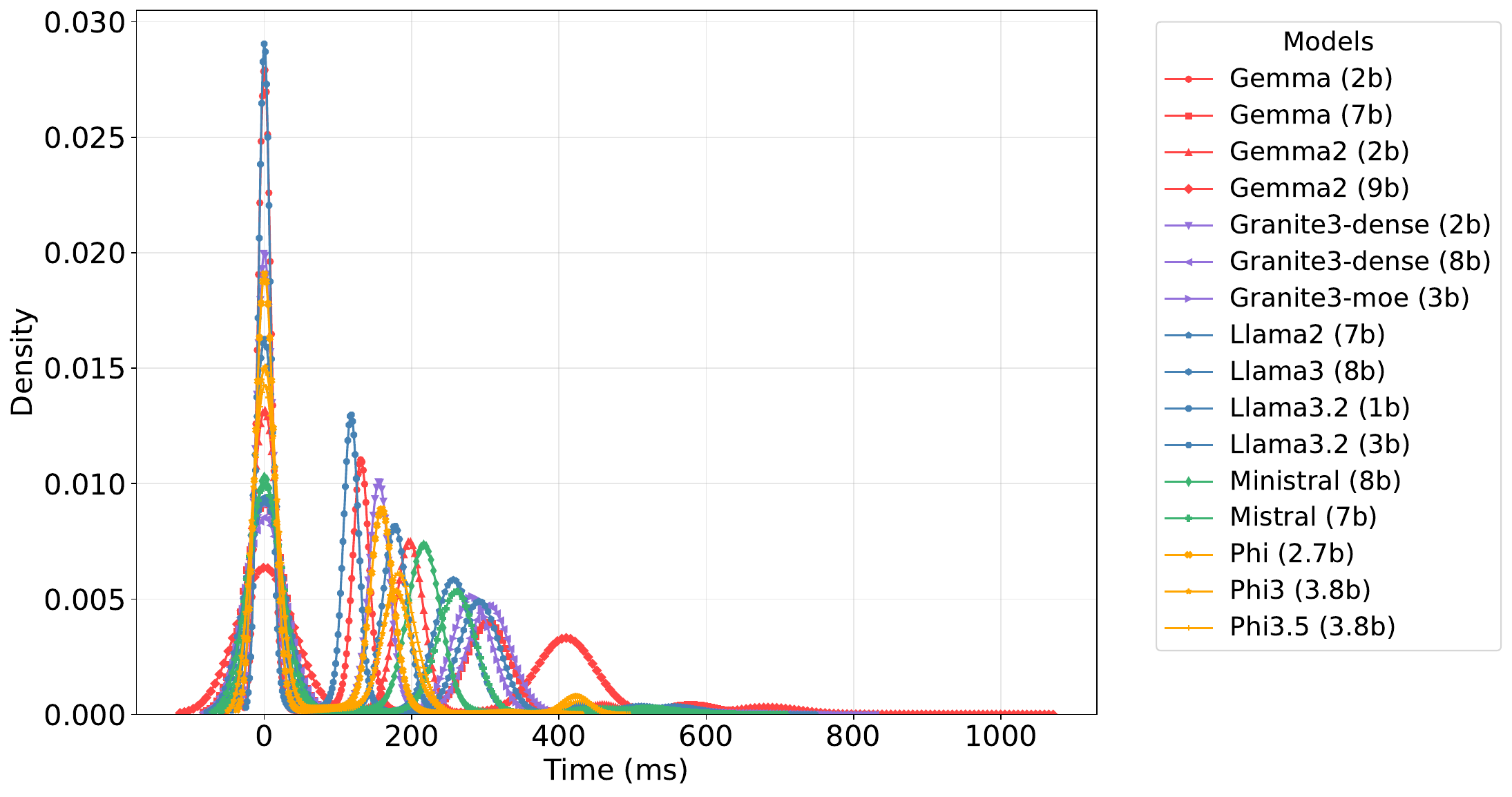}
    \caption{Probability density function of \acp{SLM} packet inter-arrival times under remote deployment conditions.}
    \label{fig:pdf_remote}
\end{figure}

In our final deployment scenario, we move from a controlled \ac{LAN} to a fully remote network where real-world network conditions and protocol overhead are introduced, including encryption, packetization, latency, jitter, routing variability, and potential congestion. Using Cloudflare's tunneling service, we enabled remote access over the internet to the Ollama server hosting the \acp{SLM}, where the client machine resides in Sweden, and the server hosting the models is located in Qatar.

Our network path analysis revealed an 8-hop route traversing multiple infrastructures, including regional ISP networks, international transit points, and Cloudflare servers, each adding a potential source of jitter, latency, and packet reordering. This setup evaluates the models' behavior under typical internet deployment conditions where users access \ac{LLM} services across geographical distances. However, this scenario represents the most challenging case for model identification, as network effects can potentially mask or distort the underlying model-specific timing patterns. Similar to the \ac{LAN} deployment, we leverage packet inter-arrival times to characterize models' behavior. 

A simple statistical analysis of the probability density distribution depicted in Fig.~\ref{fig:pdf_remote} reveals a significant overlap in packet inter-arrival time distributions among different \acp{SLM}. Compared to both the local host and \ac{LAN} scenarios, the distributions are notably broader, and the distinction between smaller and larger models is no longer visually apparent. This makes it difficult to distinguish models' identities based solely on a simple statistical analysis. This observation is quantitatively supported by Table~\ref{table1}, which shows that local host deployment models exhibit consistent performance, with mean latencies ranging from 2.910ms (LLaMA3.2 1B) to 9.935ms (Gemma2 9B) and remarkably low standard deviations (0.235-1.141ms), indicating stable token generation. These timing patterns strongly correlate with model size, as larger models consistently show higher latencies. 

On the other hand, the \ac{LAN} environment introduces minimal additional overhead, with mean latencies increased by only 0.2-0.8ms compared to the Local Host scenario, while maintaining relatively low standard deviations. In contrast, remote deployment dramatically impacts both latency and consistency, with mean latencies increased by factors of 15-20x (ranging from 57.043ms to 176.024ms), and standard deviations growing by two orders of magnitude (100-306ms). These findings indicate that real-world network conditions, such as those occurring over the internet, introduce a significant layer of complexity and variability into the timing patterns of language models. Therefore, these results suggest that simple timing-based identification methods are likely impractical for real-world environments where clients access \acp{LLM} over geographically distributed infrastructure. Reliable model identification under such conditions requires more sophisticated approaches, combining machine learning techniques with advanced feature engineering to overcome these network-induced confounding effects.

\begin{table*}[h!]
\centering
\caption{Comparison of model timing metrics across Local Host, LAN, and Remote Network environments. Mean and standard deviation values are reported for each scenario.}
\label{table1}
\begin{tabular}{|c|c|c|c|c|c|c|}
\hline
\multicolumn{1}{|c|}{\multirow{2}{*}{\textbf{Model}}} & \multicolumn{2}{c|}{\textbf{Local Host }} & \multicolumn{2}{c|}{\textbf{LAN }} & \multicolumn{2}{c|}{\textbf{Remote Network }} \\
\cline{2-7}
& \textbf{Mean (ms)} & \textbf{Std (ms)} & \textbf{Mean (ms)} & \textbf{Std (ms)} & \textbf{Mean (ms)} & \textbf{Std (ms)} \\
\hline
Gemma2:2b          & 4.757  & 0.532  & 5.073   & 5.992   & 92.491  & 140.935 \\\hline
Gemma2:9b          & 9.935  & 0.739  & 10.281  & 6.938   & 176.024 & 306.142 \\\hline
Gemma:2b           & 3.113  & 0.500  & 3.533   & 6.774   & 60.179  & 105.225 \\\hline
Gemma:7b           & 7.308  & 0.739  & 7.722   & 7.445   & 122.211 & 186.898 \\\hline
Granite3-dense:2b  & 4.264  & 0.592  & 4.536   & 5.808   & 77.297  & 124.723 \\\hline
Granite3-dense:8b  & 8.033  & 0.699  & 8.254   & 6.109   & 141.283 & 241.541 \\\hline
Granite3-moe:3b    & 7.229  & 0.235  & 7.478   & 5.443   & 130.655 & 200.566 \\\hline
Llama2:7b          & 6.197  & 0.577  & 6.388   & 5.013   & 125.487 & 186.944 \\\hline
Llama3.2:1b        & 2.910  & 0.446  & 3.203   & 5.741   & 57.043  & 100.862 \\\hline
Llama3.2:3b        & 4.353  & 0.535  & 4.682   & 6.115   & 80.970  & 125.695 \\\hline
Llama3:8b          & 7.044  & 0.628  & 7.349   & 5.957   & 124.175 & 184.154 \\\hline
Ministral:8b       & 8.612  & 0.649  & 8.678   & 5.406   & 114.694 & 158.620 \\\hline
Mistral:7b         & 6.291  & 0.603  & 6.568   & 5.547   & 115.432 & 187.360 \\\hline
Phi3.5:3.8b        & 4.416  & 1.141  & 4.833   & 6.479   & 88.051  & 137.776 \\\hline
Phi3:3.8b          & 4.223  & 0.811  & 5.045   & 9.410   & 83.558  & 134.219 \\\hline
Phi:2.7b           & 3.783  & 0.638  & 4.141   & 6.172   & 78.295  & 133.684 \\\hline
\end{tabular}
\end{table*}

\textbf{Fingerprinting \acp{SLM} Using Network Traffic and ML.}  Network traffic analysis is an active research area to manage and secure networks based on their traffic flow characteristics~\cite{papadogiannaki2021survey,azab2024network,rezaei2019deep}. Prior studies have leveraged statistical features extracted from network traffic flow, particularly focusing on packet size and inter-arrival time patterns, to classify application data~\cite{hussain2021dark,seydali2023cbs,caprolu2021cryptomining}.
\ac{ML} techniques have demonstrated superior capability in handling noisy and complex data, effectively extracting subtle and nonlinear features that are typically beyond the reach of simple statistical models. In our experiments, model-specific patterns were clearly distinguishable in local-host and \ac{LAN} deployment scenarios; however, these patterns became obscured in the remote deployment scenario due to significant network noise and variability. To overcome this issue, we explore the application of \ac{ML} techniques to learn and extract\acp{SLM} patterns from the data, even under challenging scenarios characterized by high noise and variability.

\subsubsection*{Using Network Traffic Raw Data}
\begin{figure}[!ht]
    \centering
     \includegraphics[width=\linewidth]{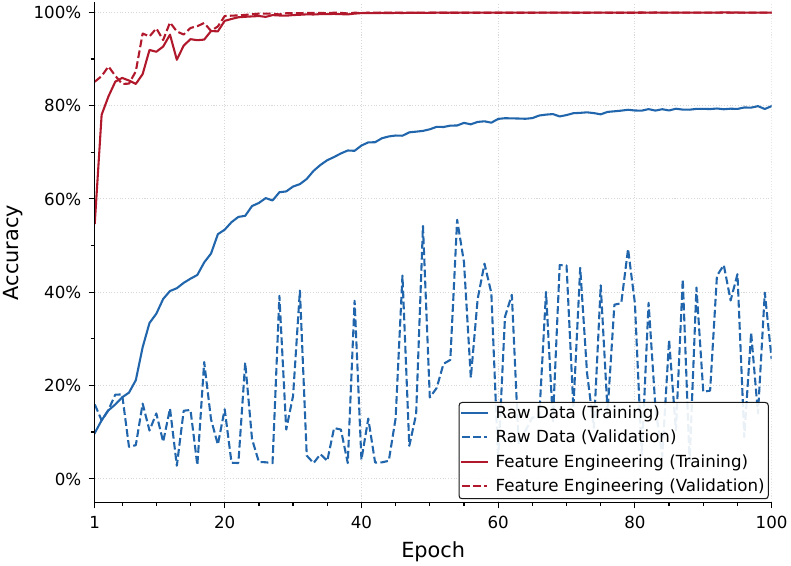}
    \caption{Training and validation accuracy curves over 100 epochs comparing model performance using raw network traffic data versus feature-engineered data.}
    \label{fig:acc_val_raw_plot}
\end{figure}
Before implementing any advanced feature engineering techniques, we first investigate whether raw network traffic data could effectively be used to identify \acp{SLM} in remote deployment scenarios. This baseline analysis aims to determine whether simple inherent patterns in raw traffic are sufficient to distinguish between different \acp{SLM}, without relying on complex pre-processing or feature extraction methods. Fig.~\ref{fig:acc_val_raw_plot} demonstrates that the \ac{DL} model is overfitting the training data and fails to generalize to the validation, which was split from the same data according to an 80/20 ratio. This suggests that the neural network model struggles with generalization, as it overfits (memorizes) noisy training data but fails to perform well on unseen samples even though they are from the same distribution~\cite{zhang2021understanding}. These findings demonstrate that raw network traffic features are insufficient for reliable \ac{SLM} fingerprinting using \ac{ML} in remote deployment scenarios. To address this limitation, a more advanced feature engineering approach is required to extract robust and discriminative characteristics from the network traffic. After applying feature engineering to raw network traffic data, the training and validation accuracies show significant improvement and convergence, as depicted by the two red curves in Fig.~\ref{fig:acc_val_raw_plot}. This convergence indicates that the model is effectively learning generalizable patterns rather than memorizing noise.

\subsubsection*{Advanced Feature Engineering}

\begin{figure}[t]
    \centering
    \includegraphics[width=\linewidth]{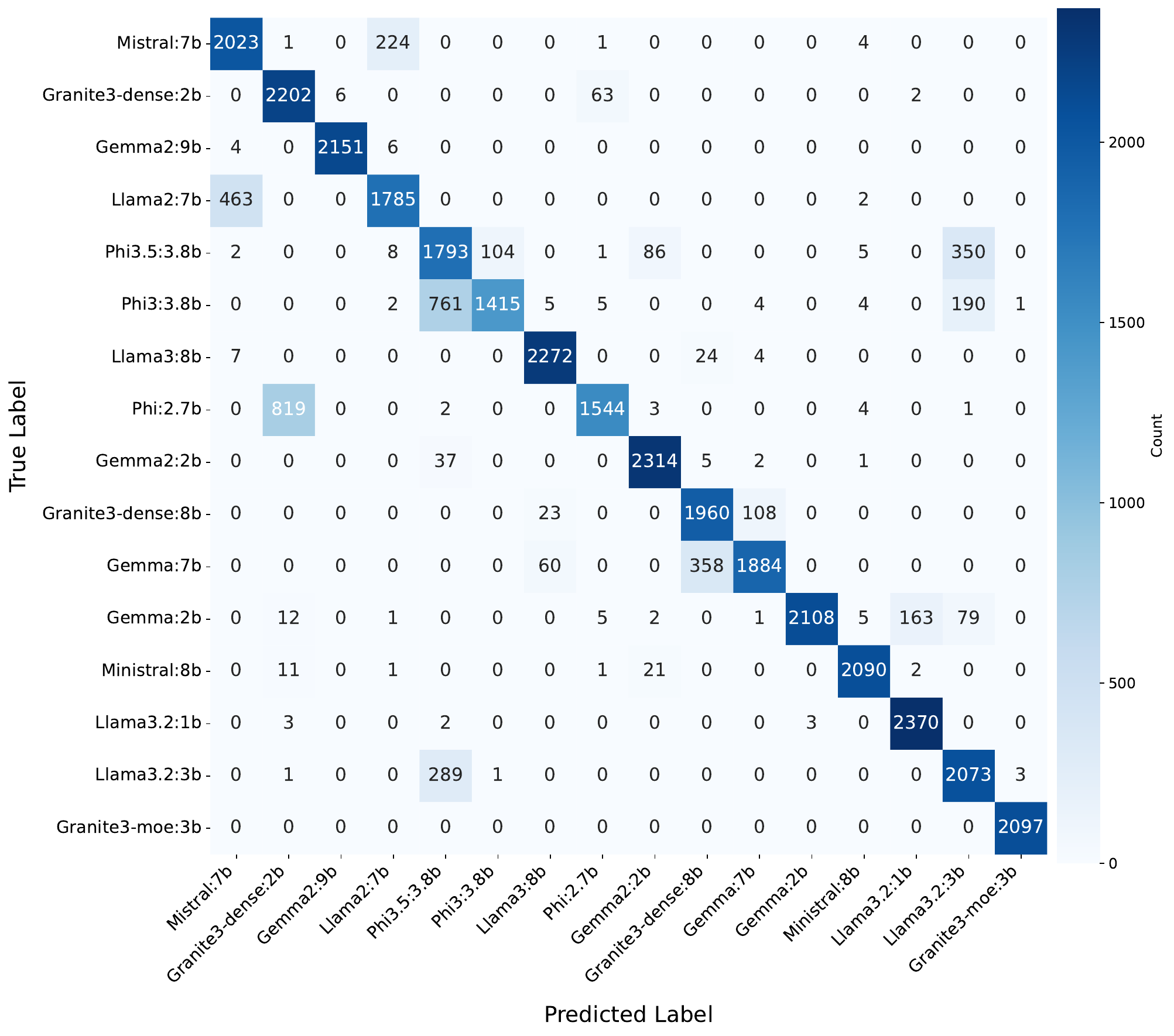}
    \caption{\acp{SLM} classification performance under remote access conditions.}
    \label{fig:confusion_matrix_slms_remote}
\end{figure}

\begin{figure}[t]
    \centering
    \includegraphics[width=\linewidth]{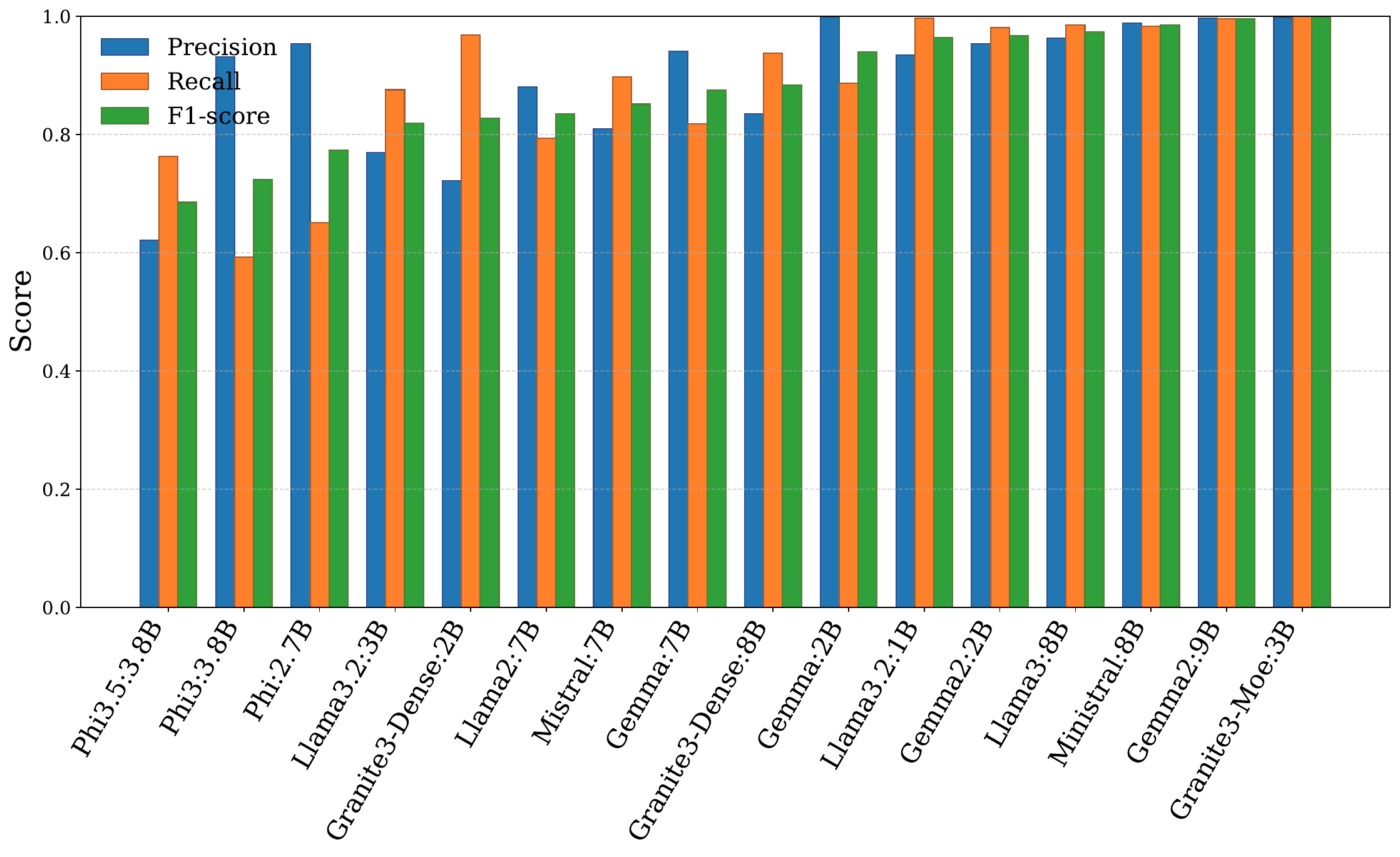}
    \caption{Precision, Recall, and F1-Score for \acp{SLM} classification under remote access conditions.}
    \label{fig:remote_fpr}
\end{figure}

Given the limitations of raw network features and the non-linear separability shown in the probability distribution, we developed a comprehensive feature engineering pipeline to capture the underlying distinctive characteristics of \ac{SLM} behavior. Specifically, our feature engineering approach transforms raw network traffic data into a rich set of discriminative higher-level statistical and temporal features designed to be robust against network-induced noise. Using a sliding window size of 0.5 seconds and step size of 0.1 seconds over the sequence of raw network traffic data, we compute 36 engineered features. These features represent a mixture of rate-based metrics (e.g., maximum average rate, burst rate), temporal statistics (inter-arrival times, entropy), correlation measures (size-time correlation), and complexity indicators (permutation entropy, entropy rate). The complete mathematical formulas of these features are provided in Appendix~\ref{sec:appendix}.

Next, we trained our model, as described in Section~\ref{alg:ModelTraining}, using the extracted data with an input size of 128, step size of 4, and batch size of 64. These parameters were selected based on iterative tuning experiments that yielded the best performance. The model was trained for 30 epochs and then evaluated on an entirely new dataset. The effectiveness of this feature engineering and training pipeline is shown in the confusion matrix in Fig.~\ref{fig:confusion_matrix_slms_remote}. The high values along the diagonal elements indicate that the classifier successfully recognizes most models. Several key observations can be drawn from the classification results. Most models are identified with high precision, with many achieving over 90\%, such as Gemma2-9B, Granite3-moe-3B, and LLaMA3-3.1B. Some confusion occurs between closely related models from the same family, such as Phi3 3.8B and Phi3.5 3.8B, which indicates similar behavior likely due to architectural similarities. Further examination of the models' performance metrics in Fig.~\ref{fig:remote_fpr} reveals that the majority of models maintain balanced precision, recall, and F1-scores above 0.85. The few exceptions occur primarily within architecturally similar model families, such as Phi models, where performance metrics drop to around 0.6-0.7. These findings confirm the robustness of the feature engineering approach and deep learning model in accurately identifying \acp{SLM} fingerprints, even when the models are hosted on remote network environments characterized by significant latency, jitter, and network-induced noise that could potentially mask the underlying token generation patterns.

\subsection{Proprietary \acp{LLM}}
Building on our successful fingerprinting of open-source \acp{SLM}, we validate our proposed solution on several popular proprietary \acp{LLM}. These models present more challenging scenarios as they are typically accessed through APIs and web interfaces, which introduces additional complexities due to server load balancing, content delivery networks, and network condition variability. We selected 10 \acp{LLM} developed by three leading companies in the field: OpenAI's ChatGPT models (ChatGPT-4, ChatGPT-4o, and ChatGPT-4o Mini), Anthropic's Claude models (Haiku, Sonnet, and Opus), and Mistral's models (Mistral-Small, Large2, Nemo, and Pixtral-Large). It is worth noting that these models are configured to stream their responses to clients, generating output in chunks (e.g., word by word or token by token) and delivering it as it is generated.  For each model, we collected approximately \emph{3,000} seconds of network traffic data for training. Following that, we trained our model using the same configuration described in Section~\ref{alg:ModelTraining}. To evaluate the robustness of our fingerprinting technique, we tested the model under three different challenging scenarios. Each represents \emph{1,000} seconds of network traffic from realistic deployment scenarios that could potentially impact the model's generalizability and detection performance:  

\textbf{Different Day.}
\begin{figure}[t]
    \centering
    \includegraphics[width=\columnwidth]{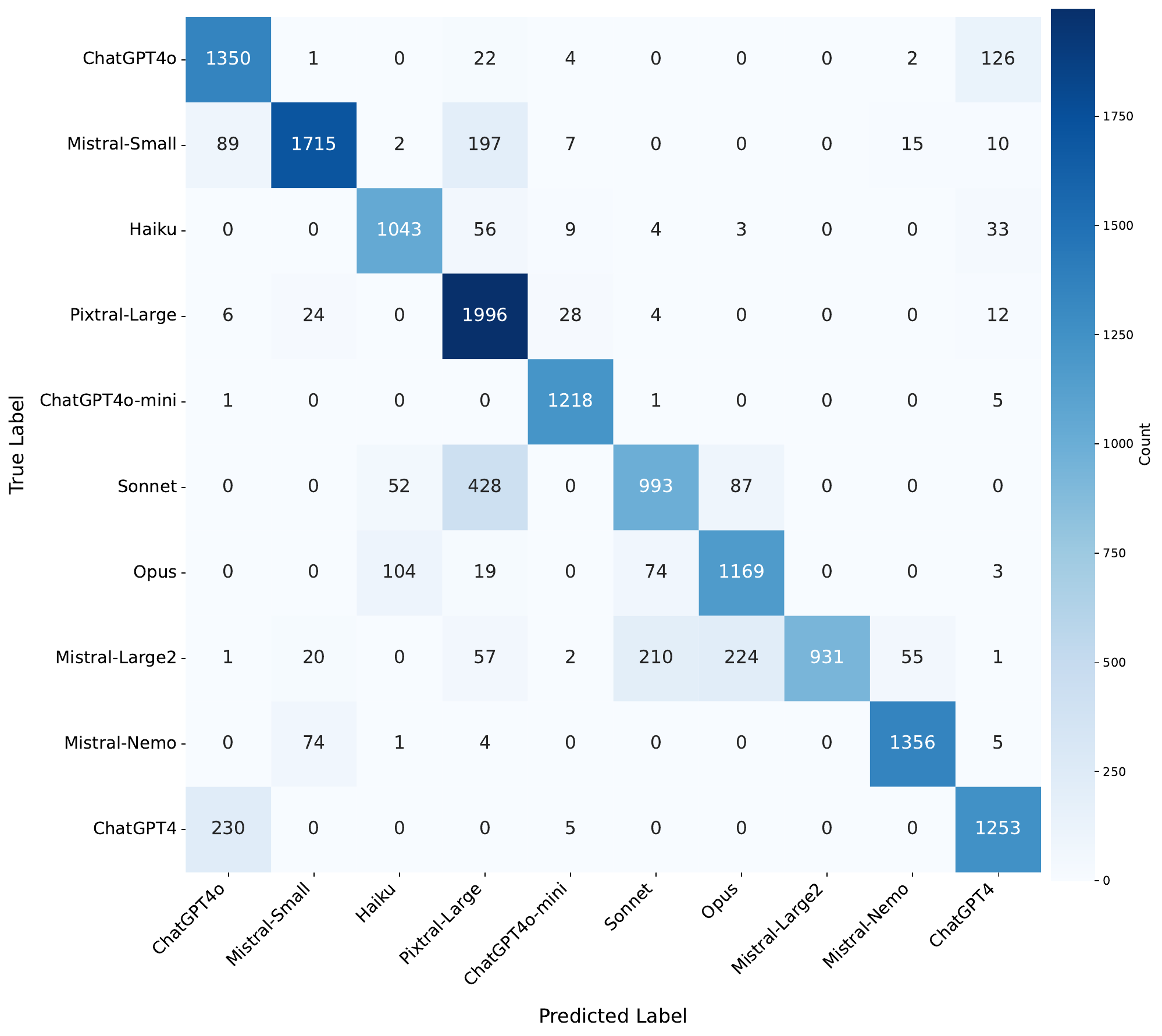}
    \caption{Model classification performance on a different day.}
    \label{fig:cm_llm_df}
\end{figure}First, we tested the model's ability to maintain its performance consistency across test data collected on a different day. This experiment aimed to determine whether temporal factors, such as varying server load and network condition variability, would significantly impact classification accuracy. Interestingly, as shown in the confusion matrix in Fig. \ref{fig:cm_llm_df}, the model maintains strong classification performance across most \acp{LLM} confirming its robustness against temporal variability. Notably, the test results reveal several insights. OpenAI's models show some inter-family confusion, particularly between ChatGPT-4o and ChatGPT-4, suggesting shared architectural characteristics. Mistral's models demonstrate robust identification with minimal cross-family confusion. On the other hand, Anthropic's models maintain clear separation from other providers. This suggests that our fingerprinting technique captures the underlying fundamental characteristics of each model's signature that remain consistent even when tested on different days.

\textbf{Different Network.}
\begin{figure}[t]
    \centering
    \includegraphics[width=\columnwidth]{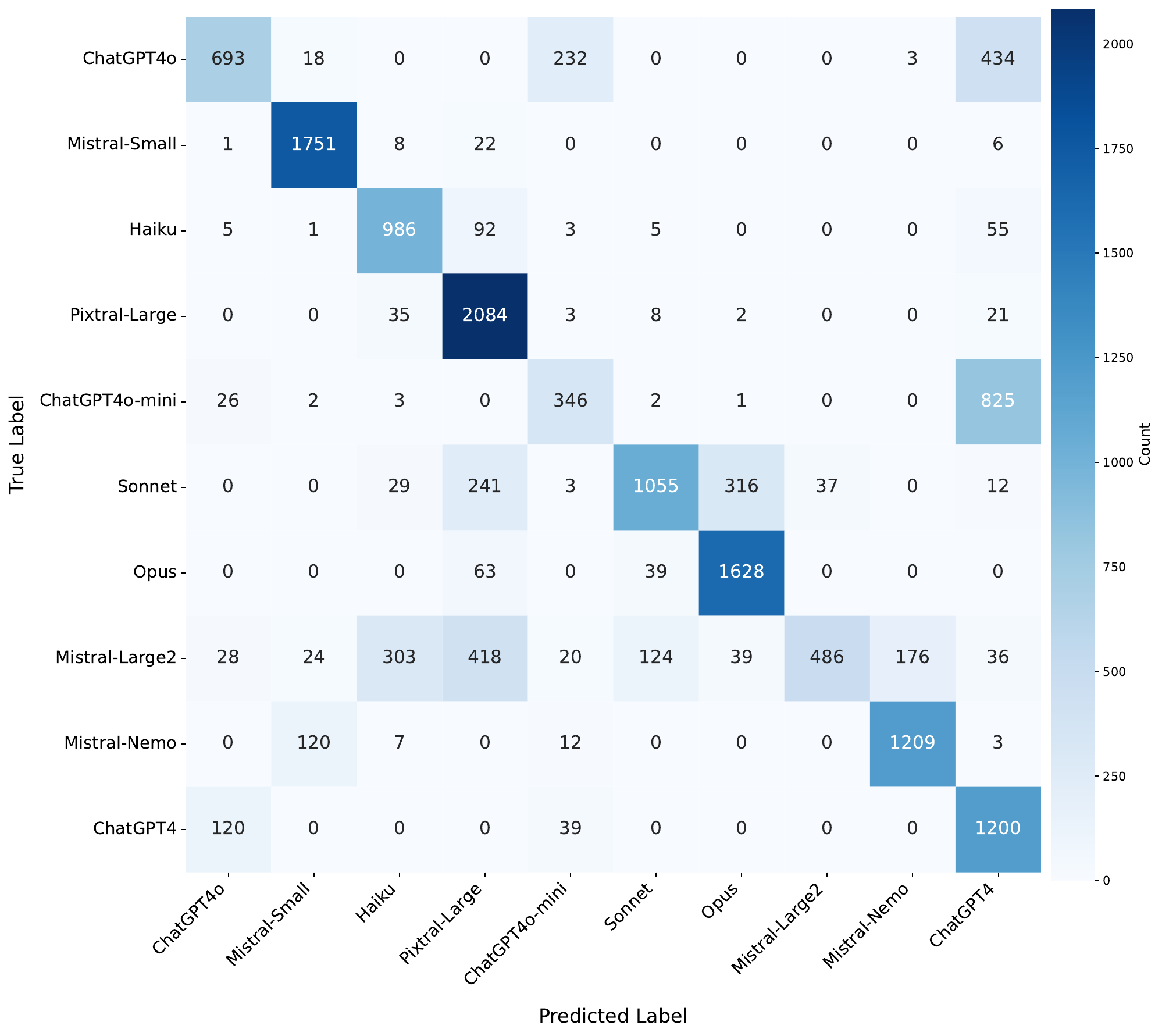}
    \caption{Cross-network classification: training on one network and testing on a different network.}
    \label{fig:cm_llm_dn}
\end{figure}Next, we tested the model on data collected from a network in different geographic locations. Specifically, while the training data was collected from a network located in Qatar, the test data was gathered from a network in Sweden. This way we can assess the model's generalizability and identify any potential bias toward the network conditions where the training was performed. Figure \ref{fig:cm_llm_dn} shows that despite the network change, most of the predicted models are correctly classified along the main diagonal indicating strong generalization across different environments. In particular, certain models such as Mistral-Small, Pixtral-Large, and Opus remain distinctive and consistently recognizable regardless of the testing environment. However, when misclassifications occur among models, they tend to be from within the same model family or from the same company.  In other words, similarities in model architecture or originating from the same company's servers may introduce some occasional confusion. Overall, these results suggest that the extracted fingerprints are not bound to a specific network setup but rather maintain their discriminative power across different network environments.

\textbf{VPN.}
\begin{figure}[t]
    \centering
    \includegraphics[width=\columnwidth]{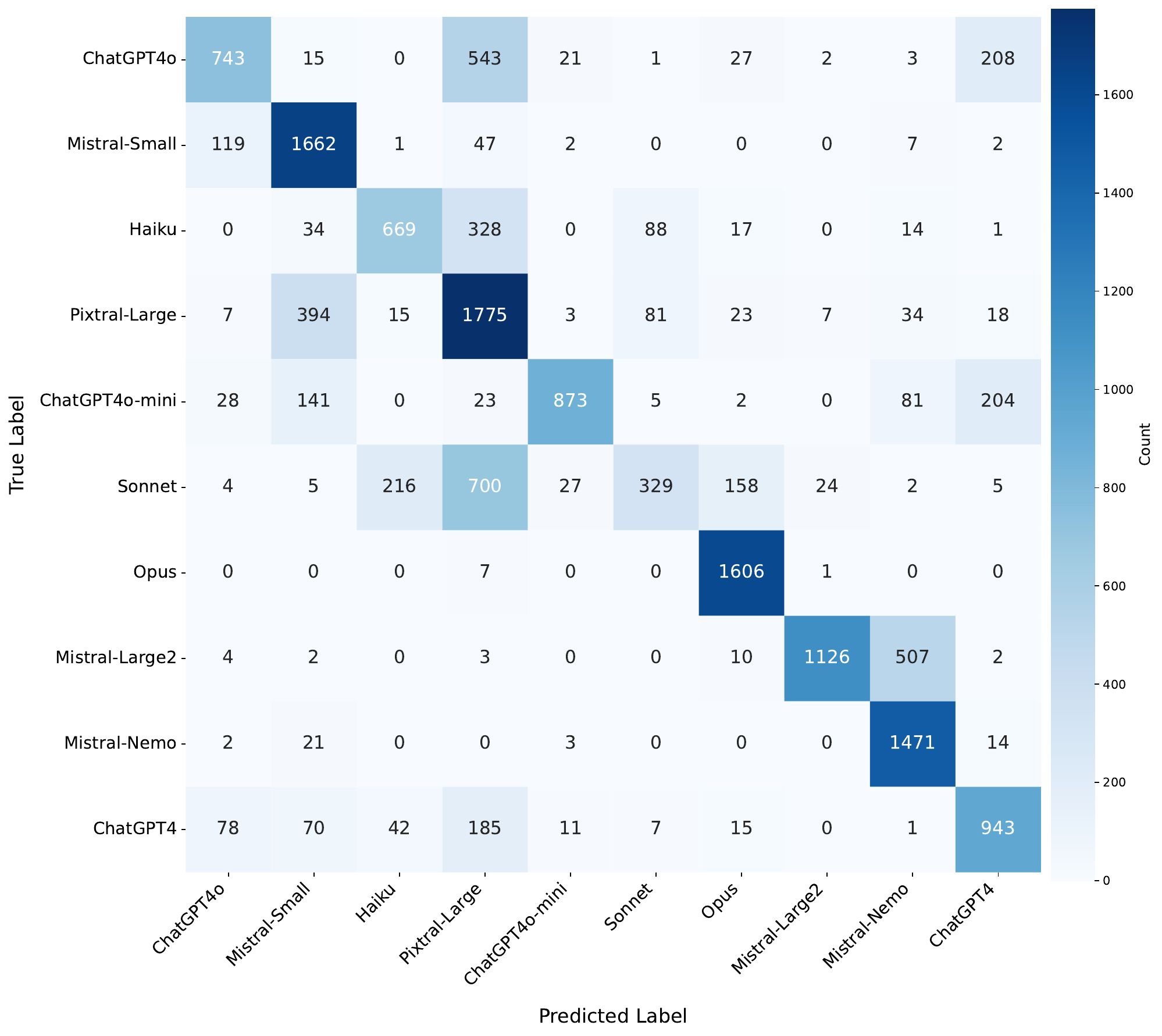}
    \caption{Model classification performance under VPN conditions.}
    \label{fig:cm_llm_vpn}
\end{figure}

Another validation scenario involves testing the model on data collected while the VPN is enabled and configured to connect to a static server in Germany. In this setup, the VPN introduces additional routing complexity and potential timing pattern distortion. We evaluate the model's resilience to this type of network obfuscation by accessing the \acp{LLM} through a VPN connection. As shown in the confusion matrix in Fig.~\ref{fig:cm_llm_vpn} the model continues to accurately recognize most \acp{LLM} correctly. However, as observed in previous tests, there is a slight increase in misclassification, particularly among models from the same vendor or family. This effect is most noticeable in Mistral models (Large2 vs. Nemo) and OpenAI models (ChatGPT-4o Mini vs. ChatGPT-4). This suggests that our fingerprinting approach remains largely effective, despite some confusion among closely related models due to VPN masking.

\begin{figure*}[!h]
    \centering
    \subfloat[]{
        \includegraphics[width=0.48\textwidth]{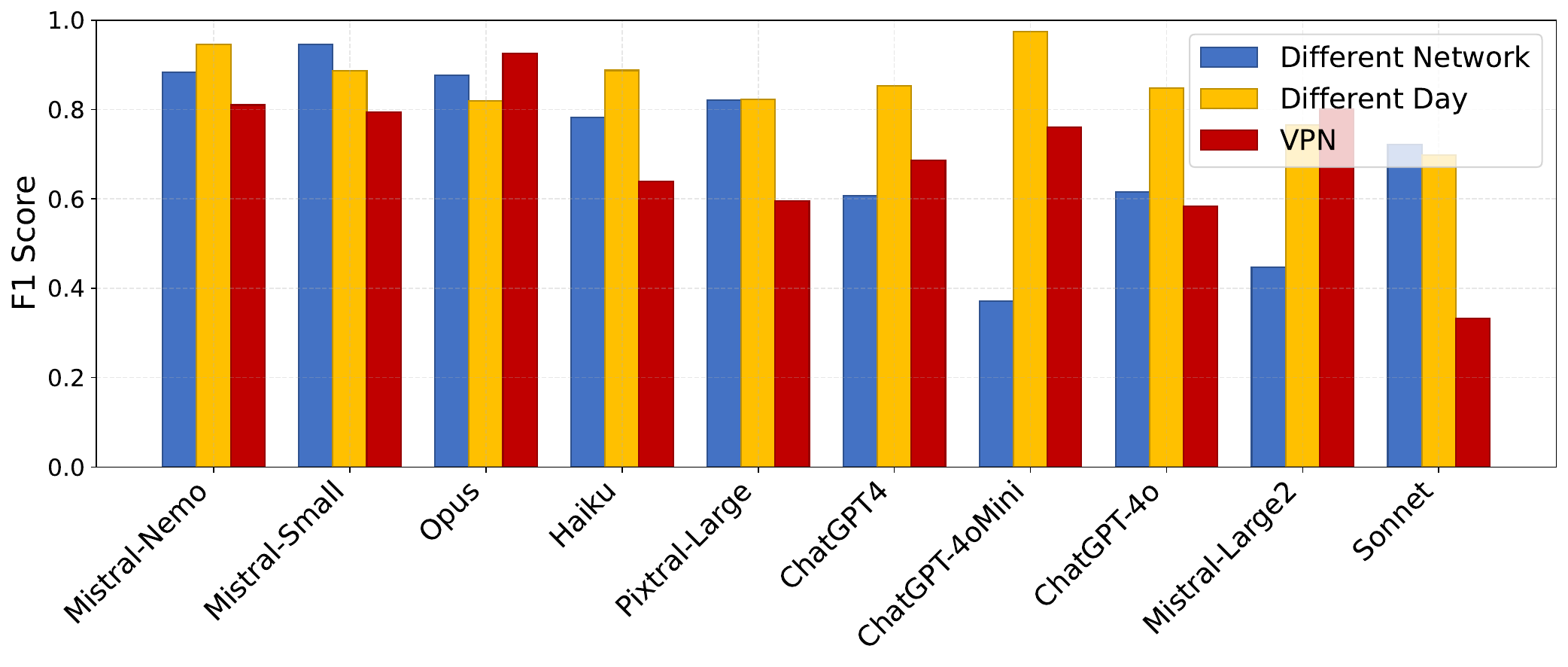}
        \label{fig:f1_comparison}
    }
    \hfill
    \subfloat[]{
        \includegraphics[width=0.48\textwidth]{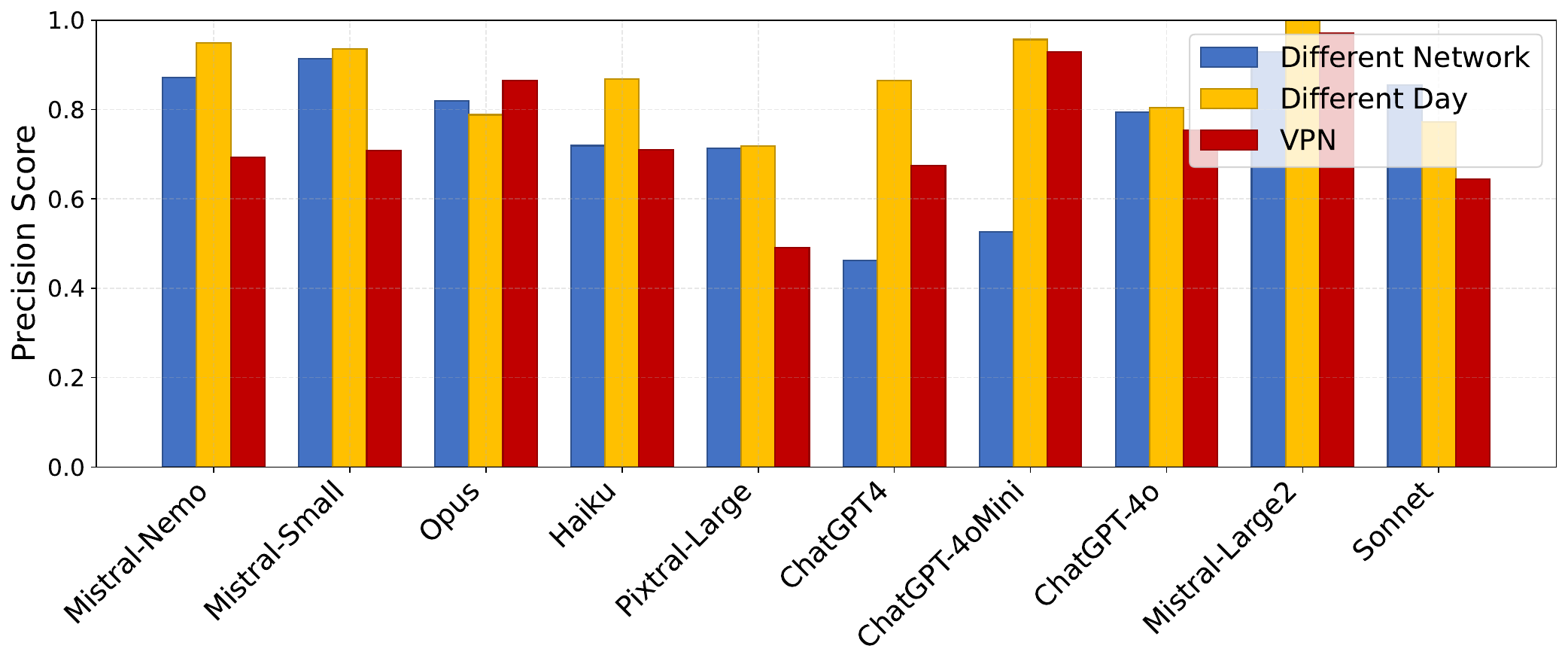}
        \label{fig:precision_comparison}
    }
    \newline
    \subfloat[]{
        \includegraphics[width=0.48\textwidth]{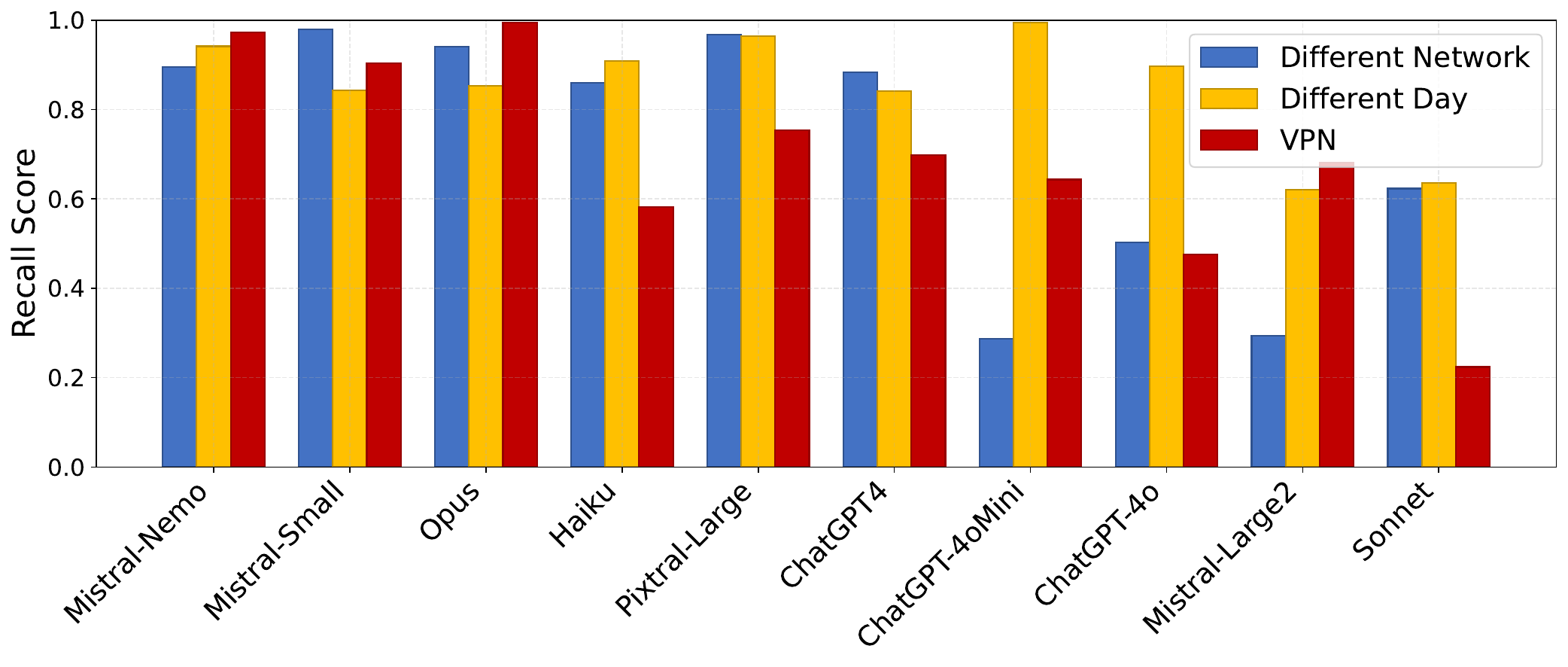}
        \label{fig:recall_comparison}
    }
    \hfill
    \subfloat[]{
        \includegraphics[width=0.48\textwidth]{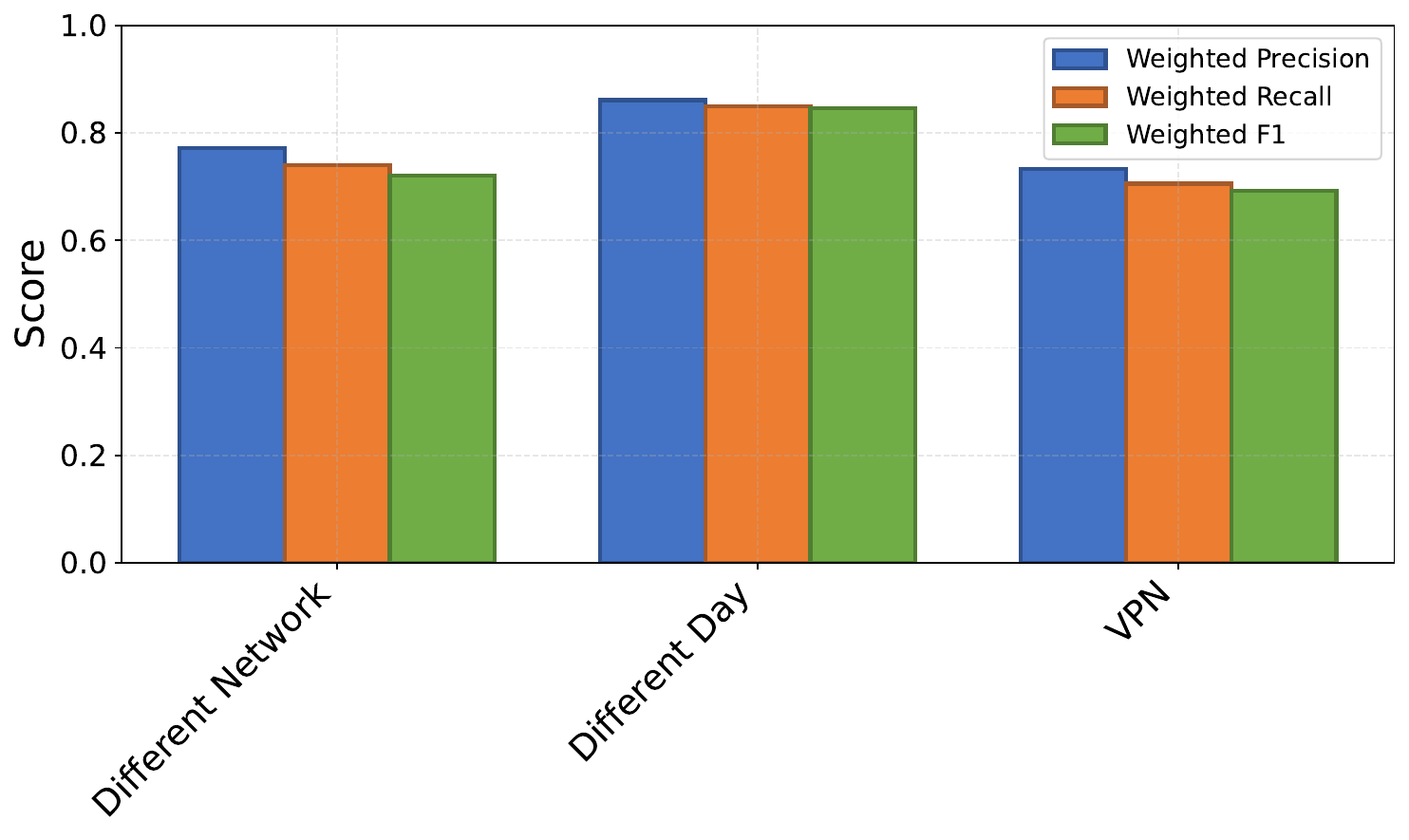}
        \label{fig:weighted_metrics}
    }
    \caption{Performance comparison across different experimental scenarios. The plots show how classification performance varies across different deployment conditions and model families, with (a) F1-score, (b) precision, (c) recall, and (d) weighted average metrics.}
    \label{fig:performance_comparison}
\end{figure*}

\textbf{Model Comparison in Different Scenarios.} As reported earlier, the overall performance of the classification model remains relatively high and robust even when tested in different conditions. To evaluate the model performance in detail using different metrics, Fig.~\ref{fig:performance_comparison} presents a comparative analysis of the model's Precision, Recall, F1-score, and weighted metrics across all three scenarios considered in our testing experiment. This comparative evaluation highlights how changing the infrastructural and obfuscation-related conditions may impact fingerprinting performance.

\subsubsection*{Different Day} The different day scenario achieves the strongest performance across all metrics. In particular, F1 scores for most models surpass 0.85, with several achieving around 0.95. Similarly, Precision and Recall follow a similar trend, suggesting that temporal variations impact fingerprint detection minimally. However, notable exceptions are Mistral-Large2 and Sonnet 3.5, where their performance across all three metrics is the lowest. This result indicates that the classifier's performance remains relatively high despite temporal variations.
\subsubsection*{Different Network} 
In the different network scenario, where the test data was collected in a different geographic location, we observe in Fig.~\ref{fig:performance_comparison} that most models experience varying degrees of resilience. In particular, models such as Mistral-Nemo, Mistral-Small, and Opus demonstrate remarkable robustness. On the contrary, models such as ChatGPT-4o Mini exhibit the most significant degradation across all metrics. In general, the results reveal that network conditions on the client side indeed affect the model fingerprint, but the degree of impact varies across models.

\subsubsection*{VPN} 
The VPN scenario represents a more challenging scenario as the model performance degrades further compared to both the different day and different network scenarios. Several models that previously showed strong performance now experience varying degrees of decline due to the additional complexity, latency, and routing obfuscation caused by the VPN. For example, models such as Mistral-Small and Mistral-Nemo managed to maintain relatively strong performance. In contrast, models such as ChatGPT-4o and Sonnet 3.5 experienced more degradation under the VPN conditions. This suggests that while VPN may make model identification more challenging, certain models retain their reliable distinct pattern even under this scenario. Overall, as shown in Fig.~\ref{fig:weighted_metrics}, the weighted metrics demonstrate that the different day scenario achieves the best performance (approximately 0.85), followed by the different network scenario (approximately 0.75), while the VPN scenario shows the lowest performance (approximately 0.7) across all weighted metrics (Precision, Recall, and F1-score).

\textbf{Feature Visualization.}

\begin{figure}[h]
    \centering
 
      \includegraphics[width=\linewidth]{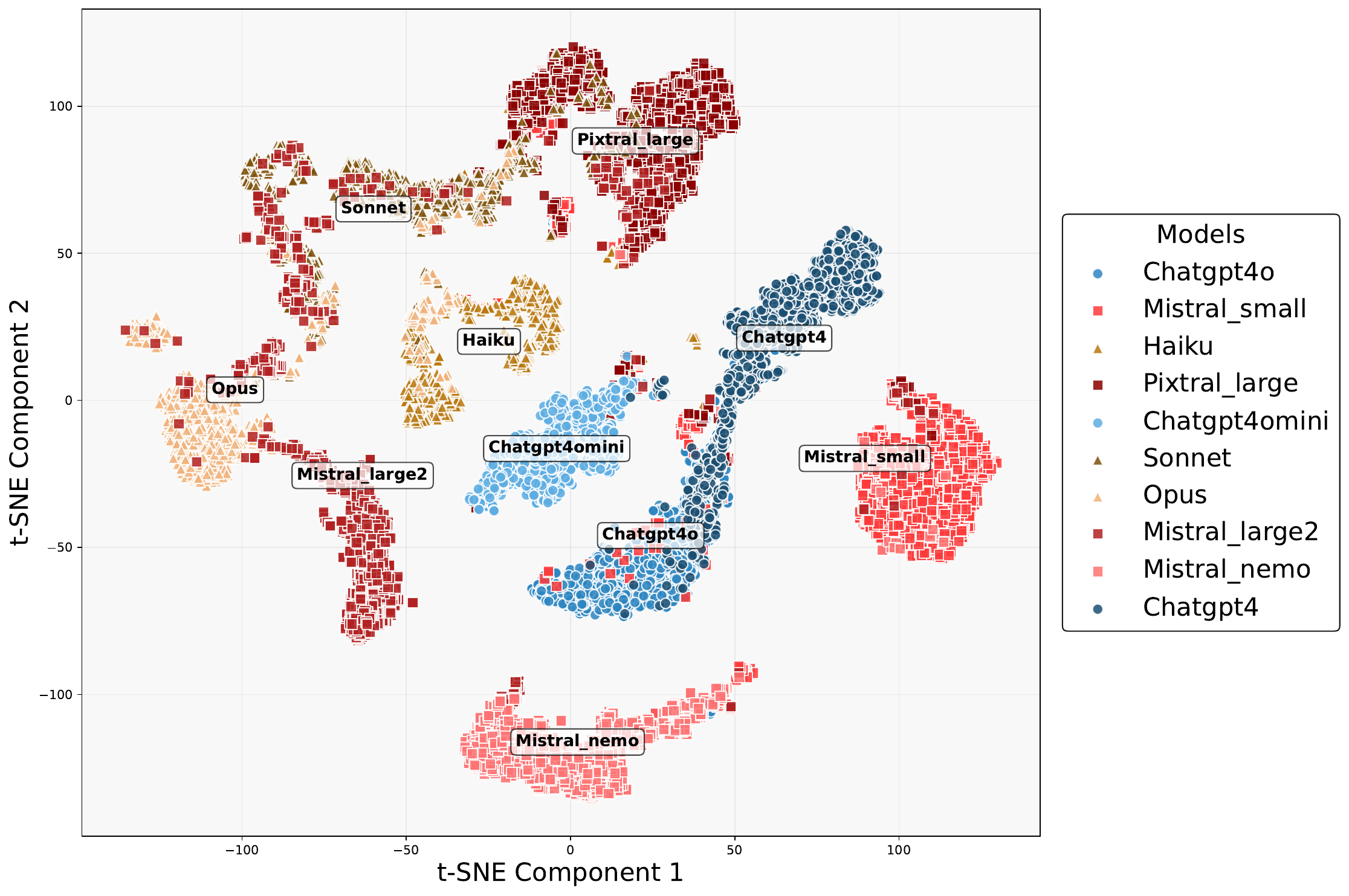}
    \caption{t-SNE 2D projection of the trained model's final-layer features, illustrating clear clustering of different \acp{LLM} fingerprints}
    \label{fig:tsne_propiertary}
\end{figure}

While the classification metrics showed the effectiveness of the model in recognizing \acp{LLM} identities, it is critical to inspect how models process and represent their distinctive characteristics internally. To visualize these high-dimensional feature representations learned by the network, we apply t-SNE dimensionality reduction to the final layer outputs, projecting them into 2D space. As shown in Fig.~\ref{fig:tsne_propiertary}, t-SNE reveals that data points belonging to the same model cluster together, forming spatially distinguishable groups. This suggests that these non-overlapping clusters are unique even after undergoing complex dimensionality reduction, indicating that these are inherent features of \acp{LLM}. Moreover, the compactness of these clusters indicates stability and consistency in the learned features, with tighter clusters suggesting lower variability in the model's behavior patterns. Furthermore, models from the same provider tend to exhibit similar behavior patterns, as shown by their close proximity in the visualization space. For example, the ChatGPT family models (ChatGPT-4, ChatGPT-4o, ChatGPT-4o Mini - shown in blue shades) cluster near each other, with particularly tight grouping between ChatGPT-4 and ChatGPT-4o. Anthropic's Opus and Haiku (brown shades) show some regions of overlap, and Sonnet appears in close proximity to other Anthropic models in certain areas. In general, this distinct separation confirms the effectiveness of our feature engineering and training pipeline in successfully capturing and distinguishing the fingerprints of the \acp{LLM}.

\section{Conclusion} 
\label{sec:conclusion}
In this paper, we demonstrated that autoregressive language models have unique temporal generation patterns that can be used for model identification even when responses are encrypted and transmitted over a network remotely. We proposed a feature engineering and training pipeline to capture the underlying model's signature and evaluated its effectiveness on both SLMs and LLMs. Our method successfully identified models with weighted F1 score 85\% on a different day, 74\% across a different network, and 71\% when accessed through a VPN, maintaining robustness despite network variability and temporal changes. This work provides a robust, real-time technique for language model identification at the network layer and enhances the security and trustworthiness of systems where language models are deployed.

\balance
\bibliographystyle{IEEEtran}
\bibliography{main}
\newpage

\appendix
\section*{Notation}

\begin{table}[h!]
\centering
\caption{Description of symbols used}
\label{table:symbols}
\begin{tabular}{|c|c|}
\hline
\textbf{Symbol} & \textbf{Description} \\ \hline
$B_i$           & Size of the $i$-th packet in bytes \\\hline
$\Delta t_i$    & Inter-arrival time between packet $i$ and $i+1$ \\\hline
$N$             & Number of packets \\\hline
$w$             & Window size \\\hline
$b$             & Burst window size \\\hline
$\bar{x}$       & Mean of $x$ \\\hline
$\sigma_x$      & Standard deviation of $x$ \\\hline
$P_k(x)$        & $k$th percentile of $x$ \\ \hline
\end{tabular}
\end{table}

\section*{36 Features Derived from Raw Network Traffic Data}
\label{sec:appendix}

\section*{Rate-Based Features}
\begin{enumerate}
\item \textbf{Maximum Average Rate}
\[
\max_{t} \, \frac{8 \cdot \sum_{i=t}^{t+w} B_i}{w}
\]

\item \textbf{Maximum Burst Rate}
\[
\max_{t} \, \frac{8 \cdot \sum_{i=t}^{t+b} B_i}{b}
\]

\item \textbf{Maximum Arrival Rate}
\[
\max_{t} \, \frac{1}{\mathrm{mean}(\Delta t)}
\]

\item \textbf{Maximum Bytes per Second}
\[
\max_{t} \sum_{i=t}^{t+1} B_i
\]

\item \textbf{Bytes per Window}
\[
\frac{\sum_{i} B_i}{w}
\]

\item \textbf{Packet Rate}
\[
\frac{N}{w}
\]
\end{enumerate}

\section*{Inter-Arrival Time (IAT) Features}
\begin{enumerate}
\item \textbf{Mean IAT}
\[
\bar{\Delta t} = \frac{1}{N-1}\sum_{i=1}^{N-1} \Delta t_i
\]

\item \textbf{Standard Deviation IAT}
\[
\sqrt{\frac{1}{N-1}\sum_{i=1}^{N-1}(\Delta t_i - \bar{\Delta t})^2}
\]

\item \textbf{Coefficient of Variation IAT}
\[
\frac{\sigma_{\Delta t}}{\bar{\Delta t}}
\]

\item \textbf{IAT 25th Percentile}
\[
P_{25}(\Delta t)
\]

\item \textbf{IAT 75th Percentile}
\[
P_{75}(\Delta t)
\]

\item \textbf{IAT 90th Percentile}
\[
P_{90}(\Delta t)
\]

\item \textbf{IAT 90th to 10th Ratio}
\[
\frac{P_{90}(\Delta t)}{P_{10}(\Delta t)}
\]

\item \textbf{Large IAT Ratio}
\[
\frac{\left|\{\Delta t_i : \Delta t_i > \bar{\Delta t}\}\right|}{N-1}
\]

\item \textbf{IAT Entropy}
\[
-\sum_{i} p_i \log p_i
\]
\end{enumerate}

\section*{Time Series Features}
\begin{enumerate}
\item \textbf{Mean Size-Time Product}
\[
\frac{1}{N-1}\sum_{i=1}^{N-1} B_i \,\Delta t_i
\]

\item \textbf{CV Size-Time Product}
\[
\frac{\sigma_{B \Delta t}}{\overline{B \Delta t}}
\]

\item \textbf{Timing Regularity}
\[
\frac{1}{1 + \sigma_{\Delta^2 t}}
\]

\item \textbf{Relative Time Pattern Entropy}
\[
-\sum_{i} p_i \log p_i
\]
\end{enumerate}

\section*{Change and Acceleration Features}
\begin{enumerate}
\item \textbf{Mean Time Change}
\[
\frac{1}{N-2}\sum_{i=1}^{N-2}(\Delta t_{i+1} - \Delta t_i)
\]

\item \textbf{Standard Deviation Time Change}
\[
\sqrt{\frac{1}{N-2}\sum_{i=1}^{N-2} (\Delta^2 t_i - \overline{\Delta^2 t})^2}
\]

\item \textbf{Mean Time Acceleration}
\[
\frac{1}{N-3}\sum_{i=1}^{N-3} (\Delta^3 t_i)
\]

\item \textbf{Standard Deviation Time Acceleration}
\[
\sqrt{\frac{1}{N-3}\sum_{i=1}^{N-3} (\Delta^3 t_i - \overline{\Delta^3 t})^2}
\]
\end{enumerate}

\section*{Correlation and Entropy Features}
\begin{enumerate}
\item \textbf{Size-Time Correlation}
\[
\frac{\mathrm{cov}(B, \Delta t)}{\sigma_B \,\sigma_{\Delta t}}
\]

\item \textbf{Time Entropy Rate}
\[
\frac{H(w)}{|w|}
\]

\item \textbf{Longest Increasing Size Sequence}
\[
\frac{\mathrm{LIS}(B)}{N}
\]

\item \textbf{Longest Increasing Time Sequence}
\[
\frac{\mathrm{LIS}(\Delta t)}{N-1}
\]

\item \textbf{Size Permutation Entropy}
\[
-\sum_{\pi} p(\pi)\log p(\pi)
\]

\item \textbf{Time Permutation Entropy}
\[
-\sum_{\pi} p(\pi)\log p(\pi)
\]
\end{enumerate}

\section*{Statistical Features}
\begin{enumerate}
\item \textbf{Time Autocorrelation}
\[
\frac{\sum_{i=1}^{N-2} (\Delta t_i - \bar{\Delta t})(\Delta t_{i+1} - \bar{\Delta t})}{\sum_{i=1}^{N-1} (\Delta t_i - \bar{\Delta t})^2}
\]

\item \textbf{IAT Skewness}
\[
\frac{\frac{1}{N-1}\sum_{i=1}^{N-1}(\Delta t_i - \bar{\Delta t})^3}{\left(\sqrt{\frac{1}{N-1}\sum_{i=1}^{N-1} (\Delta t_i - \bar{\Delta t})^2}\right)^3}
\]

\item \textbf{IAT Kurtosis}
\[
\frac{\frac{1}{N-1}\sum_{i=1}^{N-1}(\Delta t_i - \bar{\Delta t})^4}{\left(\frac{1}{N-1}\sum_{i=1}^{N-1} (\Delta t_i - \bar{\Delta t})^2\right)^2}
\]
\end{enumerate}

\section*{Rate Variability Features}
\begin{enumerate}
\item \textbf{Rate Variability}
\[
\frac{\sigma_R}{\bar{R}}
\]

\item \textbf{Peak Data Rate}
\[
\max_i \frac{8\,B_i}{\Delta t_i}
\]

\item \textbf{Burst Rate}
\[
\max_t \frac{8\sum_{i=t}^{t+b}B_i}{b}
\]

\item \textbf{Burstiness}
\[
\frac{\sigma_{\Delta t} - \bar{\Delta t}}{\sigma_{\Delta t} + \bar{\Delta t}}
\]
\end{enumerate}

\end{document}